\pdfoutput=1

\documentclass[]{JINST}

\def\pod{P$\emptyset$D}

\def\ecal{ECal}
\def\pecal{\pod-ECal}
\def\becal{barrel-ECal}
\def\dsecal{Ds-ECal}
\def\tecal{tracker-ECal}

\def\numu{$\nu_{\mu}$}
\def\nue{$\nu_e$}

\def\numutonue{$\nu_{\mu}\to\nu_{e}\,\,$}

\title{The Electromagnetic Calorimeter for the T2K Near Detector ND280}

\author{
D.\,Allan$^5$,
C.\,Andreopoulos$^5$,
C.\,Angelsen$^5$,
G.J.\,Barker$^9$,
G.\,Barr$^7$,
S.\,Bentham$^3$,
I.\,Bertram$^3$,
S.\,Boyd$^9$,
K.\,Briggs$^9$,
R.G.\,Calland$^6$
% ...
\\
\llap{$^1$}STFC, Daresbury Laboratory, Daresbury, UK\\
\llap{$^2$}Imperial College London, London, UK\\
\llap{$^3$}Lancaster University, Lancaster, UK\\
\llap{$^4$}Queen Mary, University of London, London, UK\\
\llap{$^5$}STFC, Rutherford Appleton Laboratory, Oxford, UK\\
\llap{$^6$}University of Liverpool, Liverpool, UK\\
\llap{$^7$}University of Oxford, Oxford, UK\\
\llap{$^8$}University of Sheffield, Sheffield, UK\\
\llap{$^9$}University of Warwick, Coventry, UK\\
}

\abstract{
The T2K experiment studies oscillations of an off-axis muon neutrino
beam between the J-PARC accelerator complex and the Super-Kamiokande
detector.  Special emphasis is placed on measuring the mixing angle
$\theta_{13}\,$ by observing \nue~ appearance via the sub-dominant \numutonue
oscillation and searching for CP violation
in the lepton sector.  
The experiment includes a sophisticated, off-axis, near detector, the
ND280,  situated 
280 m downstream of the neutrino production target in order to measure 
the properties of the neutrino beam and to understand better neutrino
interactions at the energy scale below a few GeV. The data collected
with the ND280 are used to study charged- 
and neutral-current neutrino interaction rates and kinematics prior to
oscillation, 
in order to reduce uncertainties in the oscillation measurements
by the far detector. 
A key element of the near detector is the ND280 electromagnetic
calorimeter (\ecal), consisting of 
active scintillator bars sandwiched between lead sheets and read out
with multi-pixel photon counters (MPPCs). The \ecal~ is vital to the
reconstruction of neutral particles, and the identification of charged
particle species. 
The \ecal~ surrounds the Pi-0 detector (\pod) and the tracking region
of the ND280, and is enclosed in the former UA1/NOMAD dipole magnet. 
 This paper describes the design, construction and assembly of the
 \ecal, as well as the materials from which it is composed.  The electronic
 and data acquisition (DAQ) systems are discussed, and 
performance of the \ecal~ modules, as deduced from measurements
with particle beams, cosmic rays, the calibration system, and T2K
 data, is described.}


\begin{document}

%\linenumbers

\section{Introduction}
Many parameters of neutrino oscillations have yet to be measured
precisely.  The Tokai-to-Kamioka (T2K) experiment is a long-baseline
neutrino oscillation experiment designed to measure several of these
parameters.  It consists of three main components: a dedicated
beamline from the proton synchrotron main ring of the Japan Proton Accelerator
Research Complex (J-PARC) that is used to produce an intense beam of
muon neutrinos; a suite of near detectors situated 280\,m downstream
of the neutrino production target (INGRID and ND280) \cite{ingrid,
  t2kExperiment} that characterize the neutrino beam before the
neutrinos change flavour; and the far detector, Super-Kamiokande
\cite{sk}, which measures the oscillated neutrino beam.  Unlike
previous accelerator-based neutrino experiments \cite{minos, k2k}, T2K
uses an off-axis configuration in which the detectors sample the
neutrino beam at an angle of 2.5$^{\circ}$ to the primary proton beam,
thus providing a narrower neutrino energy spectrum peaked at
approximately 600 MeV which is optimized for neutrino oscillation
measurements using Super-Kamiokande at a distance of 295 km downstream
and assuming the current measured value of $\Delta m^2_{32}$.
The ND280 is centred on the same off-axis angle as Super-Kamiokande in
order to sample a similar portion of the neutrino flux that will be
used to measure the oscillation parameters.  The use of a near and far
detector in this way reduces the systematic uncertainty on the
measured oscillation parameters.  The main aim of T2K is to measure
$\theta_{13}$ through the appearance of \nue~ in a \numu~ beam
\cite{t2k-nue2013}, and to improve the measurements of $\theta_{23}$
and the mass difference $\Delta m^2_{32}$ through observation of
\numu~ disappearance \cite{t2k-numu}.  With recently-measured large
values of $\theta_{13}$ \cite{dayaBay, reno}, T2K has a unique role to play
in determining whether or not there is CP violation in the lepton
sector.  T2K was the first experiment to observe indications of a
non-zero value for $\theta_{13}$ in 2011 \cite{t2k-nue}.

\begin{figure}[h]
\label{fig-nd280}
\begin{center}
\includegraphics[scale=0.80]{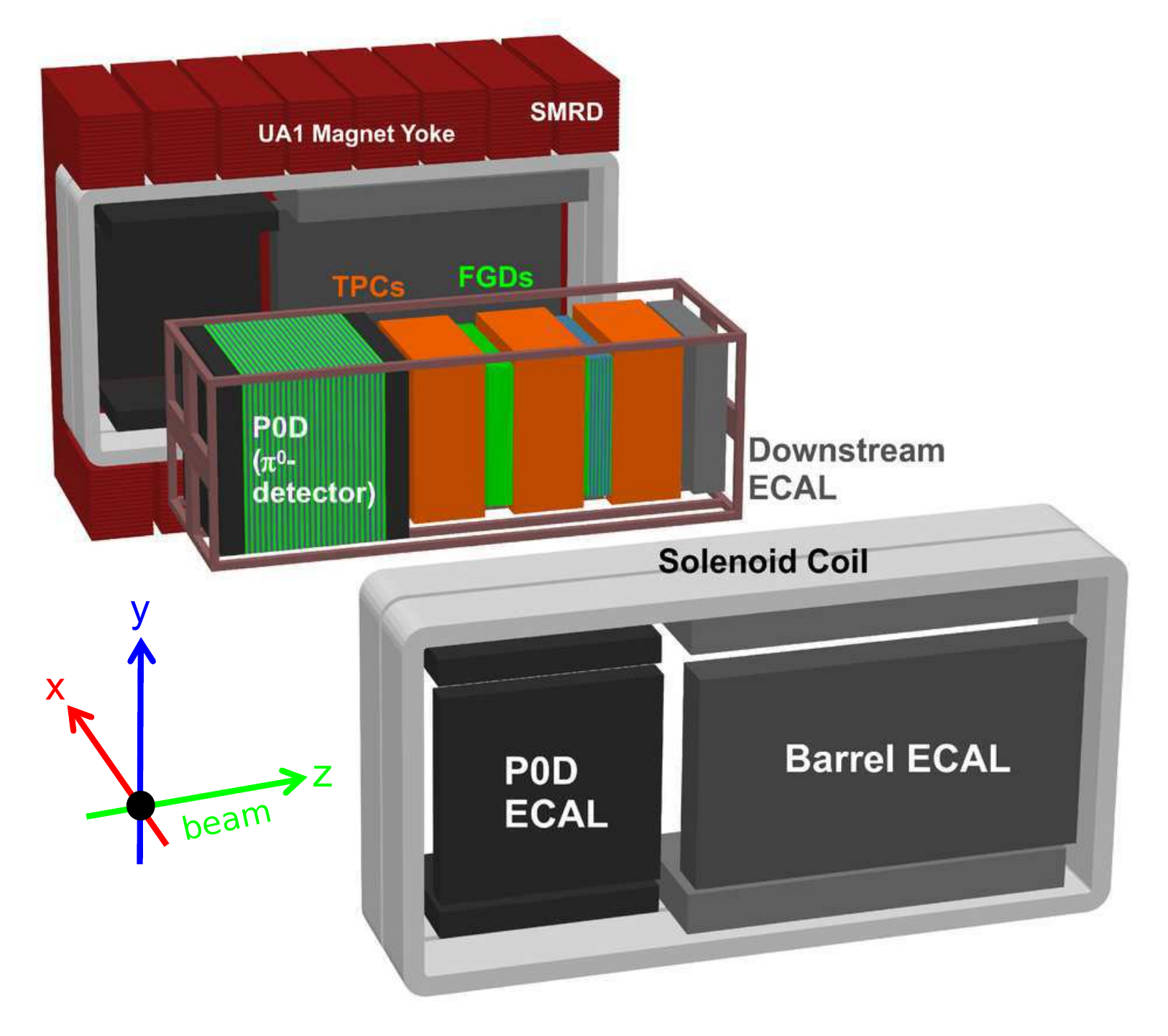}
\caption{An exploded view of the ND280 detector, showing the \pecal~ and
  \becal~ affixed to the magnet return yoke and the \dsecal~ mounted inside the basket.  The \numu~ beam enters from
  the left of the 
  figure.  The detector co-ordinate system is right-handed as shown with the
  origin at the geographical
  centre, which lies near the downstream end of the first TPC.}
\end{center}
\end{figure}

The ND280 \cite{t2kExperiment} is contained within the refurbished UA1
magnet, which in its current configuration provides a field of 0.2 T.
The detector consists of two principal sections:  the Pi-0 Detector
(\pod) \cite{pod}, optimized for identifying and measuring $\pi^0$ 
decays, and the tracker, designed for precision measurement and
identification of charged particles.  The tracker comprises three time  
projection chambers (TPCs) \cite{tpc} interspersed with two
fine-grained detectors (FGDs) \cite{fgdNIM} to provide target mass,
surrounded by an electromagnetic calorimeter (\ecal).  The \pod,
TPCs, FGDs and downstream \ecal~ (\dsecal) are mounted in a supporting
structure or `basket' which sits inside the UA1 magnet, while the
surrounding barrel- and  \pecal~ are affixed to the magnet yoke, which
splits vertically as shown in 
figure \ref{fig-nd280} to provide access for installation and
maintenance.  The yoke itself is instrumented with slabs of plastic
scintillator to act as a muon detector, the side muon range detector
(SMRD) \cite{smrd}.    With the exception of the TPCs, plastic
scintillator is used as the active material in all ND280 subdetectors.
It should be noted that, in contrast to conventional charged-particle
beams, neutrino interactions may occur at any point within the ND280 and
its immediate environment. 

The ND280 must provide a well-measured neutrino energy spectrum, flux,
and the beam neutrino composition, as well as measurements of neutrino
interaction cross-sections in order to reduce the systematic
uncertainties in the neutrino oscillation parameters.  This
information is used to predict the characteristics of the unoscillated
beam at Super-Kamiokande.  Additionally, the neutrino cross-section
measurements are 
interesting in their own right, as there are few such measurements in
the literature at present. 

The \ecal~ forms an important part of the ND280 and is essential to
obtain good measurements of neutral particles and electron/positron
showers that lead to correct particle identification and improved
energy reconstruction.  It can also be used as target material to
determine neutrino interaction cross-sections on lead.
This paper describes the design, construction and performance of the
\ecal.

\section{Overview of the calorimeter design}\label{design}

The \ecal~ is a lead-scintillator
sampling calorimeter consisting of three main regions:  the
\pecal~ which surrounds the \pod; the \becal~ which
surrounds the inner tracking detectors; and the \dsecal~ which is
located downstream of the inner detectors and occupies the last 50\,cm
of 
the basket. It is often useful to consider the \ecal~ detectors that surround
the tracker region of the ND280 together; hence the \becal~ and \dsecal~
together are referred to as the \tecal.  Altogether,
the \ecal~ consists of 13 modules: 6 \pecal~ (2 top, 2 bottom, 2 side), 6 
\becal~ (2 top, 2 bottom, 2 side), and 1 \dsecal.  The position of the \ecal~ within the ND280 is
shown in figure~\ref{fig-nd280}.  The \ecal~ modules that surround the barrel
(\becal~ and \pecal) are attached to the magnet and must have two
top and two bottom modules in order 
to allow for the opening of the ND280 magnet.

Each module consists of layers of
 scintillating polystyrene bars with cross-section 40 mm $\times$ 10
 mm bonded to lead sheets of thickness 1.75 mm (4.00 mm) in the
 \tecal~ (\pecal) which act as
 a radiator to produce electromagnetic showers and which provide a
 neutrino-interaction target.   The size of the \ecal~ is constrained
 by its position between the basket and the magnet.  A larger \ecal~
 would necessitate a smaller basket and thus smaller inner
 subdetectors.  A scintillator bar thickness of 10 mm was chosen to
 minimize the overall depth of the \ecal~ while still providing
 sufficient light to produce a reliable signal.  Scintillator bar
 widths, lead thickness and the number of layers per module were
 optimized for particle
 identification and tracking information.  Smaller bar widths are favoured
 for tracking information, and optimization studies indicated that the 
 $\pi^0$ reconstruction efficiency becomes seriously compromised for
 widths greater than 50 mm; hence, 40 mm was chosen as a compromise between
 reconstruction efficiency and channel cost.  Similarly, the lead
 thickness of 1.75 mm was chosen based upon studies of $\pi^0$
 detection efficiency. 
 The number of layers was determined by the requirement to have
 sufficient radiation lengths of material to contain electromagnetic
 showers of photons, electrons and positrons with energies up to 3
 GeV.  At least 10 electron radiation lengths, $X_0$, are required to
 ensure that more than 50\% of the energy resulting from photon
 showers initiated by a $\pi^0$ decay is contained within the
 \ecal. This requirement is satisfied in the \tecal.  More 
information about the scintillator bars and the lead are in sections
\ref{material-scintillator} and \ref{material-lead}.

The physics aims of the \tecal~ and \pecal~ are somewhat different,
and this is reflected in their design and construction.  The \tecal~
is designed 
as a tracking calorimeter to complement the charged-particle tracking 
and identification capabilities of the TPCs by providing detailed
reconstruction of electromagnetic showers.  This allows the energy of
neutral particles to be measured and assists with particle
identification in the ND280 tracker.  To this end, there are 31
scintillator-lead layers in the \becal~ and 34 layers in the \dsecal,
or approximately 10 $X_0$ and 11 $X_0$, respectively. The direction of the
scintillator bars in alternate layers is rotated by $90^{\circ}$ for
3D track and shower reconstruction purposes.  

In contrast, shower
reconstruction in the \pod~ region of the ND280 is done by the \pod~
itself, which consists of four, pre-assembled 
`Super-\pod ules', two with brass/water targets each of which
provides 2.4 (1.4) radiation lengths of material when the water is in
(out), and two with lead 
targets each of which provides 4.9 radiation lengths of material. The
role of the \pecal~ is to tag escaping energy and 
distinguish between photons and muons.  The construction of the
\pecal~ therefore differs from that of the \tecal, with coarser
sampling (six scintillator layers separated by 4 mm-thick lead sheets,
corresponding to approximately 4.3 $X_0$) and all bars running
parallel to the beam direction.  With only six scintillator layers,
the \pecal~ requires thicker lead sheets 
to ensure that photons are detected with high efficiency, that showers are
well contained, and that photon showers can be distinguished from muon 
deposits.  Simulation studies using photons and muons with energies between
65 and 1000 MeV, normally incident on a \pecal~ face, were used to
determine the optimum lead thickness.  A thickness of 4 mm was found to
provide good photon tagging efficiency ($>95\%$ for photons above 150 MeV)
and good $\mu/\gamma$ discrimination while minimizing the number of
photons that are detected only in the first layer, and might therefore be
rejected as noise \cite{thesisStill}.

Each scintillator bar has a 2~mm-diameter hole running longitudinally
through the centre of the bar for the insertion of wavelength-shifting
(WLS) fibres.  Light produced
by the passage of charged 
particles through the 
bars is collected on 1~mm-diameter WLS fibres
and transported 
to solid-state photosensors, known as multi-pixel photon counters
 (MPPCs) \cite{mppcPaper}.  The \dsecal~ WLS fibres are read out 
from both ends (double-ended readout); the \becal~ modules have a mix
of double- and single-ended readout; and the \pecal~ modules have
single-ended readout.  The fibres that are read out at one end only
are mirrored at the other end with a vacuum deposition of
aluminium. The WLS fibres and MPPCs are 
discussed more 
fully in sections \ref {material-wls} and \ref{material-mppcs},
respectively.  Each layer in each module is encased in a 20.0 mm wide $\times$ 12.5
mm high aluminium border with holes to allow the WLS fibres to exit the layer.

\begin{figure}[tbp]
\begin{center}
\includegraphics[width=\textwidth]{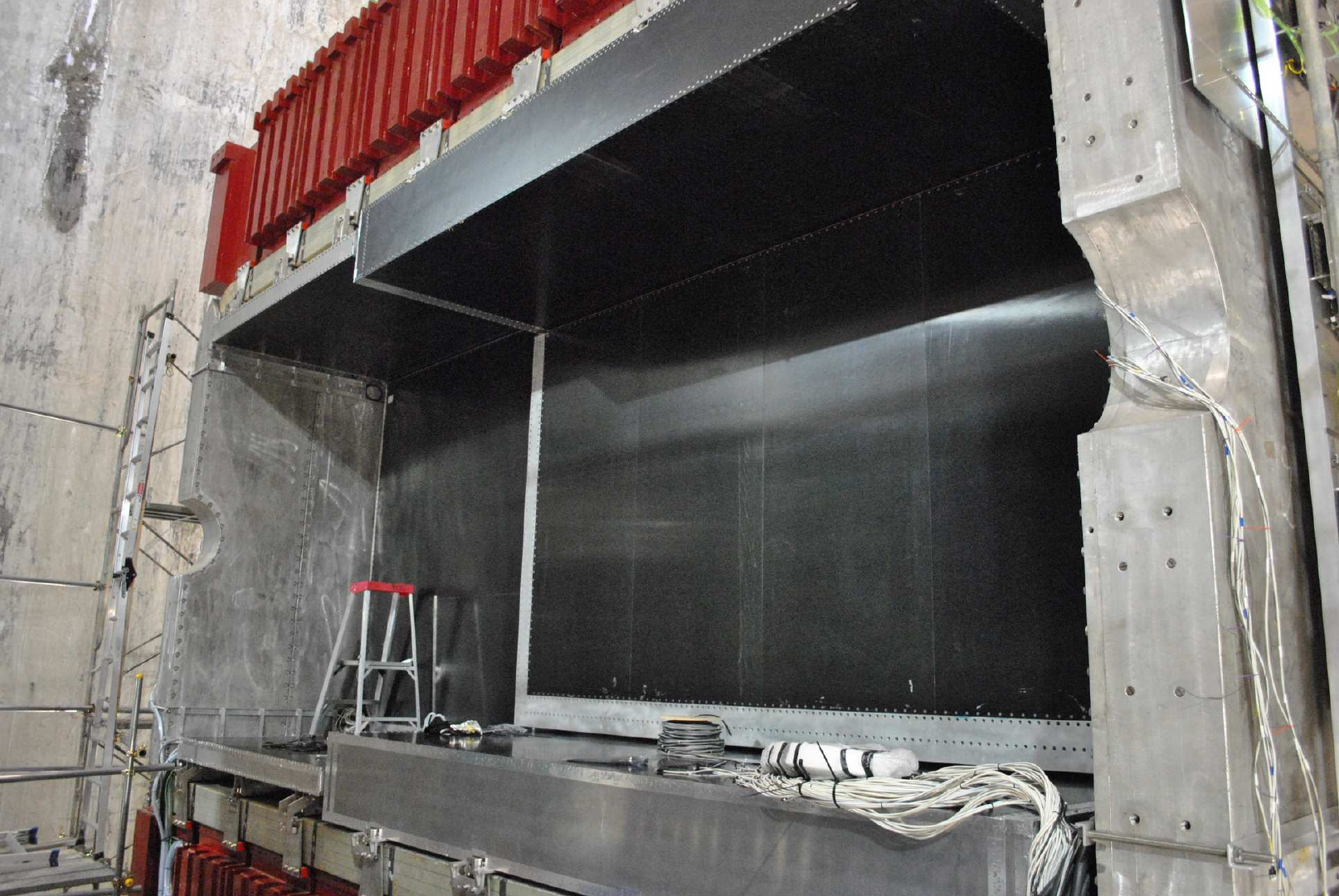}
\caption{\label{FullSideInSitu}
One entire side of the \ecal~ in situ installed in the ND280.  The
three \pecal~ modules are on the left in the figure, the three \becal~
modules are on the right. Part of the magnet yoke (top,
red) is visible.  } 
\end{center}
\end{figure}

A summary of the \ecal~ design is
shown in table \ref{tab:design}.  Further explanation is given in the
following subsections and in section \ref{construction}.  Figure
\ref{FullSideInSitu} shows one complete side of the \ecal~ in situ.
The \pecal~ is on the left and the \becal~ is on the right in the
figure.  Visible are the top, side and bottom modules for each.  Notice
that the \pecal~ is thinner than the \becal~ as described above.

\begin{table}
\begin{center}
\caption{\label{tab:design} Summary of the \ecal~design showing the
  overall dimensions, numbers of layers, length and orientation of
  scintillator bars, numbers of bars, and lead thickness for each module.}
\begin{tabular}{|l|l|l|l|}
\hline
        & DS-Ecal  &Barrel ECal & P0D ECal  \\ \hline
Length (mm)& 2300      & 4140      & 2454   \\ 
Width (mm)  & 2300      & 1676  top/bottom       & 1584  top/bottom \\ 
        &          & 2500 side             & 2898 side  \\ 
Depth (mm)  & 500     & 462                     & 155    \\
Weight (kg) & 6500 & 8000 top/bottom  & 1500 top/bottom  \\
            &      & 10000 side       & 3000 side        \\ \hline
Num. of layers &  34      &  31        &  6   \\ \hline
Bar orientation & $x/y$ &Longitudinal and Perpendicular  & Longitudinal \\ \hline
Num. of bars    & 1700     & 2280 Longitudinal top/bottom & 912 Longitudinal top/bottom    \\ 
        &          & 1710 Longitudinal sides & 828 Longitudinal sides \\     
        &          & 6144 Perp top/bottom &  \\ 
        &          & 3072 Perp sides      & \\ \hline  
Bars per layer & 50 & 38 Longitudinal top/bottom & 38 Longitudinal top/bottom\\
               &    & 57 Longitudinal side & 69 Longitudinal sides \\
               &    & 96 Perp top/bottom/sides & \\  \hline
Bar length (mm) & 2000  & 3840  Longitudinal & 2340 Longitudinal \\ 
           &      & 1520 Perp top/bottom & \\ 
           &      & 2280 Perp sides      & \\ \hline
Pb thickness (mm) & 1.75  & 1.75   & 4.0   \\ \hline

\end{tabular}
\end{center}
\end{table}

\subsection{The Downstream \ecal}\label{dsecal}
\begin{figure}[tbp]
\begin{center}
\includegraphics[width=\textwidth]{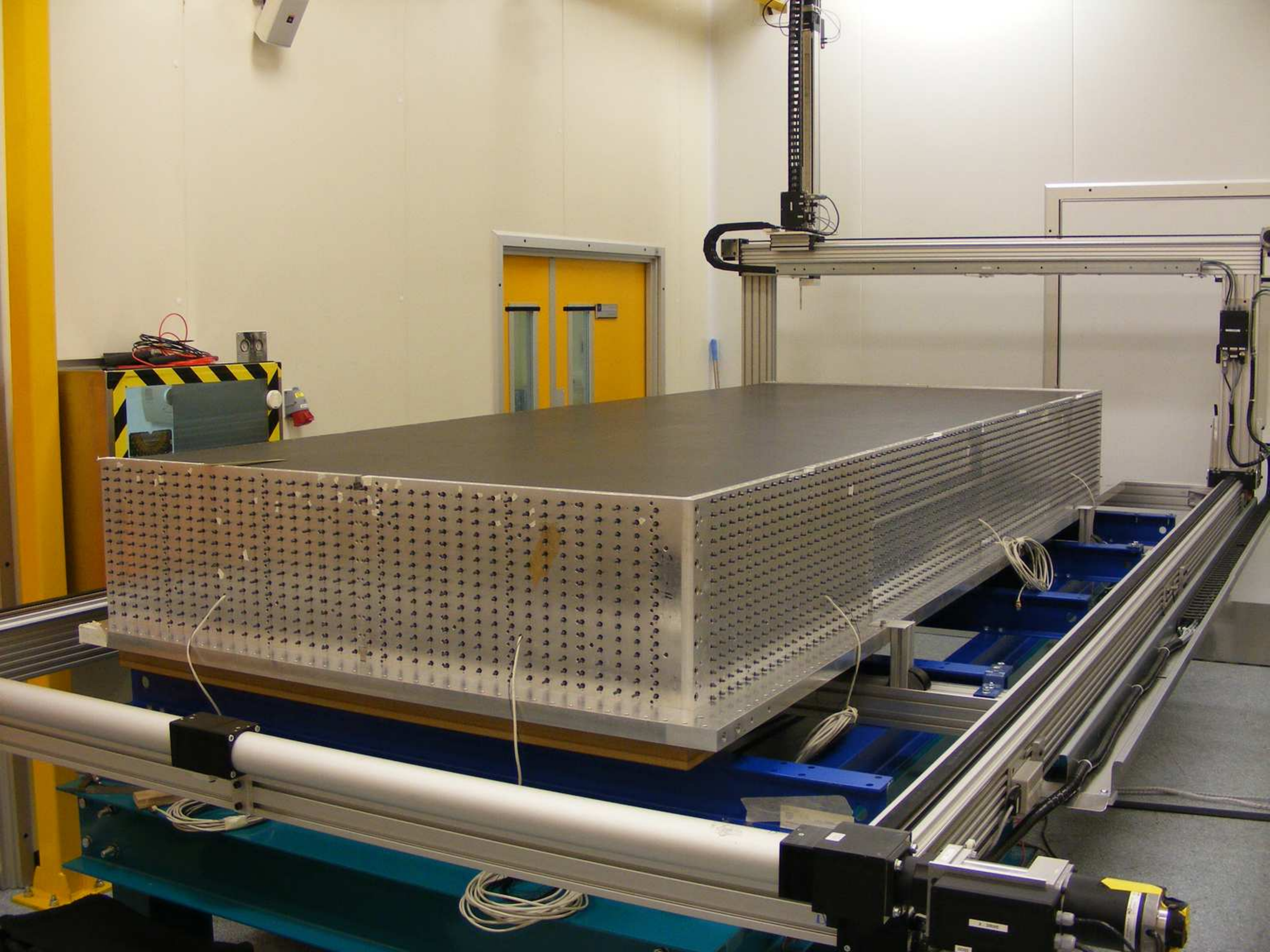}
\caption{\label{fig-topBarrel}
One of the top \becal~ 1.5 m $\times$ 4 m modules lying horizontally
during construction. The fibre ends encased in their ferrules are
visible protruding from the module bulkheads.  The structure of the 2D
scanner can be seen surrounding the module.}
\end{center}
\end{figure}

The first detector to be constructed and commissioned was the \dsecal,
which also acted as a prototype.  The outer dimensions of the \dsecal~
are 2300 mm high $\times$ 2300 mm wide $\times$ 500 mm long (depth in
the beam direction).  Each of the 34 layers has 50 scintillator bars
of length 2000~mm.  The bars of the most-upstream layer run in the
$x$-direction (horizontally) when the module is installed in the ND280
basket.  Surrounding the 34 layers on all four sides are 25~mm-thick
aluminium bulkheads, which have holes for the WLS fibres to exit.
Once outside the bulkhead, each end of every fibre is secured inside a
custom-made Teflon ferrule as discussed in section \ref{ferrules} and
shown in figure~\ref{Connector}, which is then covered by a matching
sheath that allows the WLS fibre to make contact with the protective,
transparent, resin coating of the MPPC.  This contact is maintained by
a sponge-like spring situated behind the MPPC that can absorb the
effects of thermal expansion and contraction in the WLS fibres.  The
sheath also contains a simple printed circuit board which couples
the MPPC to a mini-coaxial cable that carries the information between
the MPPC and the front-end electronic cards.  Figure
\ref{fig-topBarrel} shows the top \becal~ module during
construction.  The fibre ends in the ferrules are visible protruding
from the module bulkheads.  The electronics are
described in section \ref{electronics}.  The ferrule is designed to
latch into the sheath which is then screwed to the bulkhead in order
to hold the ferrule and WLS fibre in place, and to assure that the
coupling between the WLS fibre and the MPPC is secure.  The MPPC-WLS
fibre coupling is described in section \ref{material-wls}.

There is a 1 cm gap between the layers and the bulkheads on all sides
to leave space for the light injection (LI) system.  The LI system,
described in section \ref{li}, uses LED pulsers to deliver short
flashes of light through the gap to illuminate all of the WLS 
fibres, allowing integrity and calibration checks to be performed. 

\begin{figure}[tbp]
\begin{center}
\includegraphics[scale=1.00]{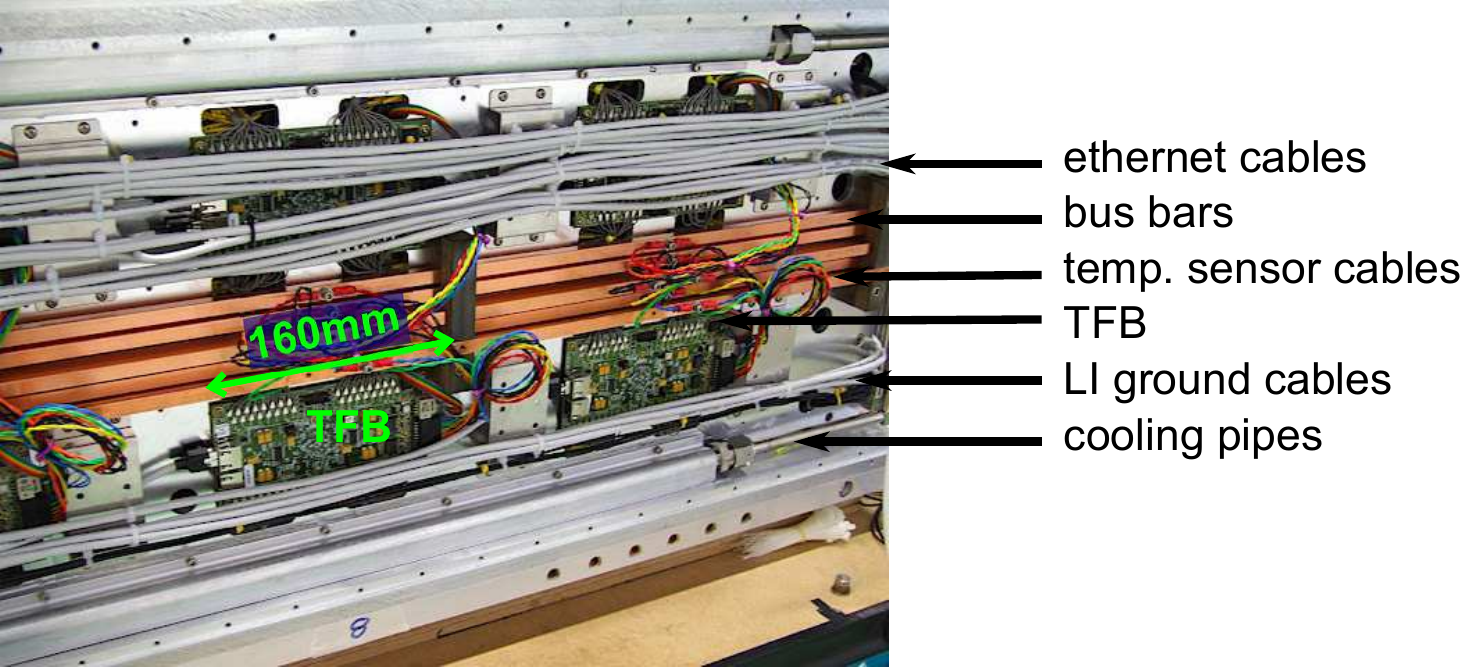}
\caption{\label{fig-dsside}
The left side of the \dsecal~ lying horizontally during construction.
When upright in situ, the bottom in the figure becomes the
upstream surface of the \dsecal~ nearest to the inner detectors, and
the top in the figure becomes the downstream surface nearest to the
magnet coils.
Shown at the top and bottom are the Cat~5e cables (commonly used as
ethernet cables), the LI cables (small, black cables at the bottom), 
  the TFBs (green cards), the external
  temperature-sensor cables 
  (multi-coloured), the cooling pipes (aluminium, top and bottom), and
  the bus bars (brass, centre) mounted on the
  cooling panels.  Air holes in the cooling panel
  are visible on the right side of the
  figure.  The Cat~5e cables at the top include both the signal and
  trigger cables; those at the bottom are signal cables.}
\end{center}
\end{figure}

Cooling panels for temperature control are located outside the
bulkheads as shown in figure~\ref{fig-dsside}.  Pipes carrying chilled
water maintain these panels at a 
constant temperature of approximately 21$^{\circ}$C; the bottom panel
also has perforated air pipes through which dry air is pumped to
prevent condensation within the module.  Large air holes through the
cooling panels and the bulkheads allow the air to flush through
the active region of the detector and escape from the module.

The Trip-T
front-end electronic boards (TFBs) are mounted on the cooling panels
using screws and thermally-conducting
epoxy resin; slots in the panels allow the cables from the MPPCs to
pass through and  terminate on the
TFBs. Each TFB has 64 channels to read out MPPCs, a built-in
internal temperature sensor, and a port that connects to an external
temperature  
sensor mounted on the bulkhead near the MPPCs in order to 
monitor the MPPC temperatures.  There are 14 TFBs per side.
Figure~\ref{fig-dsside} shows the left-side cooling panel with the
TFBs 
installed.   

The cooling panels are protected by anodized aluminium cover panels,
while the 2000 mm $\times$ 2000 mm outer surfaces are covered by
carbon-fibre panels in order to minimize the mass of these dead
regions.  These cover panels form the outside of the module.  Each
carbon-fibre panel consists of two sheets of carbon-fibre of
dimensions 2059 mm wide $\times$ 2059 mm long $\times$ 1.2 mm thick.
A foam layer of 22.6 mm thickness is sandwiched between the two
sheets, making the entire panel 25 mm deep.  The carbon-fibre sheets
are glued tongue-in-groove into an aluminium border that is 120 mm
wide and 25 mm thick, making the dimensions of the entire
carbon-fibre-aluminium panel 2299 mm $\times$ 2299 mm $\times$ 25 mm
thick.  In situ, the \dsecal~ sits upright inside the basket.  Water,
dry air and high-voltage enter through the bottom cover panel.  The
information to and from the TFBs is carried by shielded Cat~5e
cables which exit through a patch panel in the bottom cover panel.  In
addition to these 56 signal cables, there are 28 trigger cables,
routed through 28 cable glands in the top cover panel of the module, which
come from the TFBs that are connected to the MPPCs located near the
downstream edge of the module.  Data from these channels form part of
the ND280 cosmic ray trigger.  Upon exiting the cover panels, the
signal Cat~5e cables are connected to the readout merger modules
(RMMs) which are attached to the outer surface of the cover panels.
The RMMs are discussed in section \ref{electronics}.  The trigger
cables are connected to fan-in cards on the top of the module.  The
\dsecal~ is the only \ecal~ module which forms part of the ND280
cosmic ray trigger system.

\subsection{The Barrel \ecal}\label{becal}

The four \becal~ top and bottom modules are 4140 mm long (parallel to
the beam) $\times$ 1676
mm wide $\times$ 462 mm high, with 31 lead-scintillator
layers:  16 (including the innermost layer) with 1520~mm-long scintillator bars
running perpendicular to the beam direction, and 15 with 3840~mm-long 
bars running 
longitudinally, i.e. parallel to the beam direction. 

The structure of each of the \ecal~ modules is very similar to that
described above for the \dsecal, except that the perpendicular bars
have single-ended readout, with the fibres mirrored on the end that
is not read out.  The
mirrored ends of the fibres terminate just inside the scintillator
bars, whereas the ends that are read out exit through the bulkhead
as in the \dsecal.  The longitudinal
bars all have double-ended readout.

In order to minimize the non-active gap in the \ecal~ running down
the centre of 
the ND280 between the two top or the two bottom modules, the mirrored
ends of the fibres in the perpendicular bars are in the centre of the
ND280 and the readout ends are at the sides.  This made it possible to
replace the thick 
aluminium bulkhead, cooling panels and cover panels that form the
structure on the other three sides of each module with a thin
aluminium cover panel, allowing the two top and the two bottom
modules to be placed closer together and minimizing the dead material
between them.

The two side \becal~ modules are 4140 mm long $\times$ 2500 mm wide
$\times$ 462 mm deep, with 31 lead-scintillator
layers:  16 (including the innermost layer) with 2280-mm-long scintillator bars
running perpendicular to the beam direction, and 15 with 3840-mm-long 
bars running  
longitudinally, i.e. parallel to the beam direction.   As in the top
and bottom \becal, the perpendicular bars are single-ended readout,
with the mirrored ends of the fibres at the top and the readout ends
at the bottom of the modules.  

Unlike the \dsecal~ which has carbon-fibre panels
on both the most upstream and downstream faces, the \becal~ modules have a
carbon-fibre panel on the innermost face, but the outermost face
has an aluminium panel which provides the required structure for
attaching the module to the magnet yoke.

\subsection{The \pod~ \ecal}\label{pecal}

The most noticeable difference between the \tecal~ modules and the
\pecal~ modules is the smaller size of the \pecal, which has six 
scintillator-lead layers and is only 155 mm deep. The four top/bottom
modules are 1584 mm wide, the two side modules are 2898 mm wide, and
all six are 2454 mm long. The \pecal~ also required thicker 4 mm lead
sheets as a consequence of having fewer layers. All of the
scintillator bars, 38 in each top/bottom module layer and 69 in the
side layers, are oriented parallel to the beam, read out on the upstream end and mirrored on the downstream end. The smaller size also allowed simplifications to be made in the construction.

\begin{figure}[tbp]
\begin{center}
\includegraphics[scale=1.00]{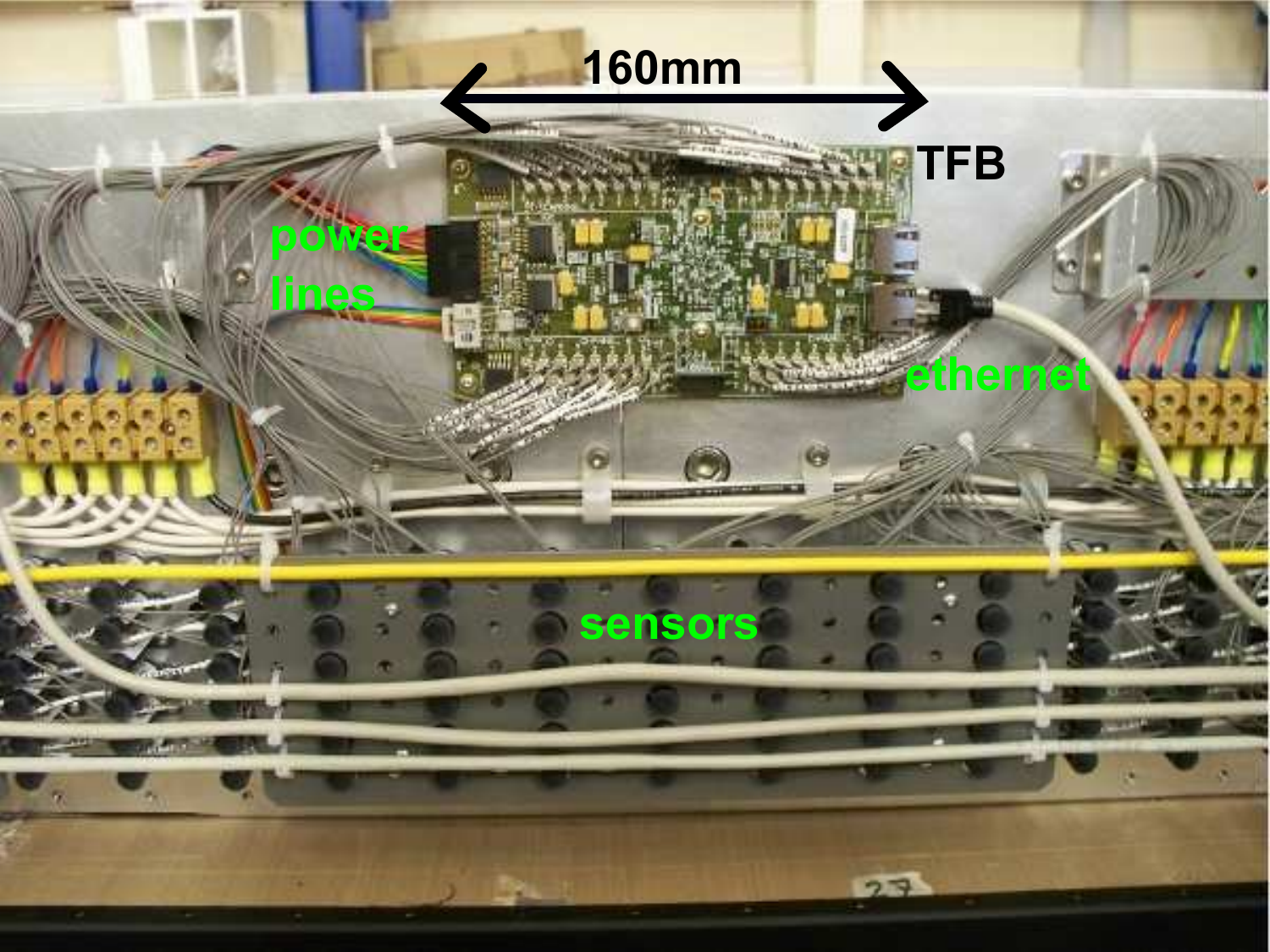}
\caption{\label{fig-podecal}
Closeup on the readout side of the upstream bulkhead for a \pecal~
side module. The black cases at the bottom house the MPPCs
(photosensors) coupled to 
the optical fibres where they emerge from the inner detector. Each
MPPC is connected via a thin grey cable to the TFB mounted above, which
receives its power from the coloured cables on its left and is read
via the Cat~5e (ethernet) cable on its right.} 
\end{center}
\end{figure}

While the readout electronics of the \tecal~ detectors were mounted on
separate cooling plates, the \pecal~ TFBs are attached directly to the
upstream bulkhead as shown in figure~\ref{fig-podecal}. This bulkhead
extends 100 mm above the height of 
the main detector box so that the TFBs can be mounted next to the
region where the optical fibres emerge. The water cooling pipes are
then recessed into grooves running along the exposed face behind,
through which the dry air system also passes. Because there are no
more than seven TFBs in any one module, the boards could have their
power provided via standard copper cables instead of the bus bars used
in the \tecal. The readout region was then protected by an anodized
aluminium cover. 

Structurally, the \pecal~ modules used the same carbon-fibre panels as
the \tecal~ on the bases, but all had solid aluminium bulkheads to
form the lids. Onto this were bolted the structures required to mount
the detectors on the near detector magnet: cast aluminium rails for
the side modules, and roller cages for the top/bottom modules. 

\section{Materials}
\subsection{Scintillator bars}\label{material-scintillator}
The \ecal~ scintillator bars were made at the Fermi National
Accelerator Laboratory (FNAL) from extruded polystyrene doped with
organic fluors at concentrations of 1\% PPO and 0.03\% POPOP. The
polystyrene scintillator bars have a $0.25 \pm 0.13$ mm
coating of polystyrene co-extruded with TiO$_2$ providing light
reflection and isolation.   The
scintillator was chosen to have the same composition as that used for
the MINOS detectors \cite{MINOSscintillator}.  Each bar has a
cross-section of ($40.0^{+0.0}_{-0.4}$) mm wide $\times$
($10.0^{+0.0}_{-0.4}$) mm deep with a $2.0 \pm 0.2$ mm hole running
down the centre for the insertion of 1 mm-diameter WLS fibre, which is
discussed in more detail in section \ref{material-wls}.  Each bar was
cut to the appropriate length during the quality assurance (QA)
process described below, to within 0.1 mm.  The number of bars of each
length is shown in table \ref{tab:design}.  Including the 10\% extra
that were made to replace any rejected during the QA process, there
were a total of more than 18,300 bars shipped from FNAL.

The scintillator bars underwent both mechanical and optical QA tests.
The frequency of testing was reduced in some 
cases as the extrusion process at FNAL was refined due to feedback
from the QA group.  During the mechanical QA 
tests,  the sizes of 100\% of the bars were 
checked for width and thickness using
custom-made Go and No-Go gauges that could slide easily along bars if
the sizes were within tolerance, and a visual inspection was made for
flatness and  squareness. The bars 
were then cut to length.  The hole position and diameter 
were checked using digital callipers for 100\% of the bars for the
\dsecal, and for 10\% of the bars subsequently.  Optical QA was
carried out on 10\% of the bars for the \dsecal~ and 5\% of the bars
subsequently. 

For the first shipments of bars, which were used in the \dsecal, the
hole diameter 
was found to vary from 1.75 mm to 3.50 mm and the 
shape was typically elliptical rather than round.  This did not affect
the light yield of the bar; however, it did have 
consequences for the layer construction.  If the hole was too small,
the locator pins used to hold the bar in place while the epoxy cured
could not be inserted into the bar; if the hole was too large, glue
would enter the hole around the edges of the locator pin and block
the subsequent insertion of the WLS fibre.  Approximately 10\% of
\dsecal~ bars were rejected due to this problem.  Subsequent shipments
of bars for use in the other \ecal~ modules did not have this problem.

Optical QA was carried out on the scintillator bars in order to ensure
a consistency of response to minimum-ionizing particles (MIPs).  To
this end, a cosmic ray (CR) telescope was constructed consisting of a
light-tight enclosure for the scintillator bars and their
photo-readout, a triple-coincidence trigger, and a simple data
acquisition (DAQ) system.  The bar being tested did not form part of
the trigger \cite{thesisDavies}.

The CR telescope consisted of three 4 cm $\times$ 6 cm
scintillator pads coupled to 2-inch photomultiplier tubes (PMTs) which
were biased to
approximately 2 kV each.  The coincidence trigger consisted of a
simultaneous trigger from one
scintillator pad above and two pads below the bar being
tested.  This criterion excluded showers and random coincidences
from the PMTs and reduced the selection of CRs to those that
entered the telescope at a small angle from the zenith,
eliminating the need to correct for path-length differences in the bar
due to
differing angles of incidence.  The CR telescope registered an average
of 450 triggers per hour as expected from calculations.

Each bar to be tested for light yield was first wrapped in two layers
of microfibre blackout material as a form of flexible dark box.  In
order to obtain a robust signal over the background electronic noise,
the light signal of the test scintillator was collected by three
1~mm-diameter WLS fibres, rather than one, in the central hole of the
scintillator bar, which was possible due to the hole diameters being
larger than specified, as discussed above.  The three fibres were
coupled to the centre of a 
2-inch PMT with optical grease to
improve light transmission.  The PMT was biased at a voltage of 2.55
kV.

\begin{figure}[htpb]
\begin{center}
\includegraphics[scale=0.50]{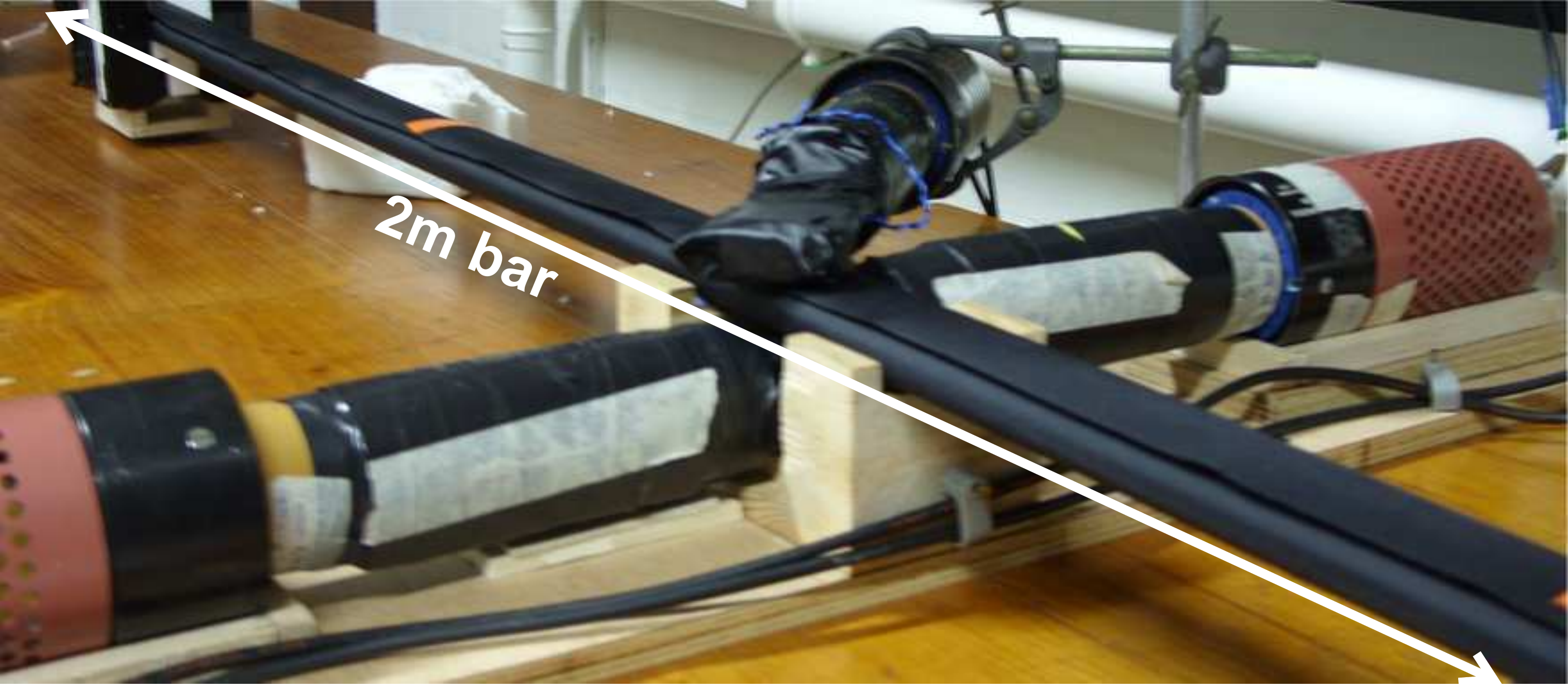}
\caption{\label{fig-QASetup}
The CR telescope showing the 2 m-long scintillator bar in light-tight
microfibre sleeve (centre) and the triple-coincidence trigger.  The trigger
scintillator pads are situated with one above and two below the test
bar.}
\end{center}
\end{figure}

The bar readout end was confined inside a series of plastic adaptors
to enable easy replacement of the bars for testing, and to ensure a
consistent, reproducible optical coupling to the PMT.  A viewing port
was built to enable visual verification of the coupling between the WLS
fibres and the PMT, and to act as an input port for an LED-based
light-injection system that was used to calibrate the single
photo-electron (PE) peak in the PMT.  The optical QA setup is shown in
figure~\ref{fig-QASetup}.

Output signals from the trigger PMTs were fed into a into a computer
running LabVIEW that 
produced  histograms of  the charge integrals and
exported the data for subsequent
analysis \cite{thesisDavies}.  A baseline was established using 12
scintillator bars from 
an early delivery.  After analysis and calibration using the single PE peak,
the light yield for these 2 m bars read out at a position 66 cm from
the readout end was found to be $34.2 \pm 0.6$
PE/MIP with $\sigma \approx 5$ PE/MIP.  Bars with a light yield within
$2\sigma$ of the baseline light yield 
were accepted.  This criterion did not reject any bars.

\subsection{Lead}\label{material-lead}
The target mass and radiator for each
layer is provided by a thin sheet of lead stiffened with
$2.0 \pm 0.2$\% antimony.  Traces of other
metals are below 0.15\%. 
Each sheet of lead was coated with black, quick-drying,
metal-conditioning primer, CELEROL-Reaktionsgrund 918 (Reaction
Primer), before being used in a layer, in order to 
protect personnel from the harmful effects of lead and prevent any
possibility of leaching between the lead and the scintillator 
which might degrade the light-yield qualities of the scintillator over
time.  The lead thickness for all layers in the \becal~ and the \dsecal~
is $1.75 \pm 0.10$~mm, and for the \pecal~ is $4.0 \pm 0.3$~mm.  The
tolerances were determined by the manufacturer and are due to
the lead fabrication process.  The
widths and lengths differ depending upon the size of the module.  Due
to the technical difficulty of producing large sheets of thin lead,
for the \dsecal~ and \becal~ each layer of lead includes more than a
single sheet.   

The \dsecal~
sheets upon delivery were $1008 ^{+4}_{-0}\,\, \mbox{mm wide} \times 2019
\pm 4 \,\, \mbox{mm long}$ and subsequently were cut to lengths of $2016
\pm 1$ mm during the layer construction.  Two sheets were laid side by
side to make up a 
single \dsecal~ layer.   

The \becal~ top and bottom modules have lead dimensions $765
^{+4}_{-0}\,\, \mbox{mm wide} \times 3858 ^{+4}_{-0} \,\, \mbox{mm
  long}$.  Two sheets were laid side by side to make up a single
\becal~ top or bottom layer.  The \becal~ side modules have lead
dimensions $2330^{+4}_{-0}\,\, \mbox{mm wide} \times 964.5^{+4}_{-0}
\,\, \mbox{mm long}$, with the total length being provided by laying
four sheets along the length of each layer.

The \pecal~ top and bottom modules have lead dimensions $1528
^{+4}_{-0}\,\, \mbox{mm wide} \times 2356^{+4}_{-0} \,\, \mbox{mm
  long}$ and the side modules have lead dimensions $2770
^{+4}_{-0}\,\, \mbox{mm wide} \times 2356^{+4}_{-0} \,\, \mbox{mm
  long}$.
 Unlike for the \becal~ and \dsecal~ layers, due to the thickness of these
sheets it was possible to 
produce sheets that were wide enough for the entire layer.

\subsection{Wavelength-shifting fibre}\label{material-wls}

All \ecal~ modules used Kuraray WLS fibres of the same type:
Y-11(200)M, CS-35J, which are multi-clad fibres with 200 ppm WLS dye.
The fibre diameter 
was specified to be $1.00^{+ 0.02}_{- 0.03}$ mm.  The fibres were
delivered for processing as straight `canes' to the Thin Film Coatings
facility in Lab 7 at FNAL. All fibres were cut to length with a
tolerance of $\pm 0.5$ mm and both ends were polished in an
ice-polishing process where a batch of around 200 fibres at a time
(800 fibres per day) 
were diamond-polished using ice as a mechanical support. The \pecal~
fibres and shorter fibres from the \becal~ modules were then
`mirrored' on one end in batches of 800 -- 1000 using an aluminium sputtering vacuum process,
maximizing the amount of light available to be read out from the
opposite end of the fibre and saving on double-ended readout. A thin
layer of epoxy was applied to each mirrored end for
protection. Table~\ref{TAB-FIBREORDER} summarizes the full \ecal~ WLS
fibre consignment. In total, over 17,000 fibres were processed at Lab
7 and passed through the QA process described below. This represented
about a $10\%$ contingency over the total number of fibres needed to
complete construction of the \ecal.
\begin{table}
\begin{center}
	\caption{\label{TAB-FIBREORDER}The consignment of WLS fibres used
          in the \ecal~ construction.} 
\begin{tabular}{|l|l|l|l|} \hline
\ecal~ Fibre Type & Length (mm) & Quantity & Processing \\ \hline
\becal~ Side     & 2343        &  3072 & cut, ice polish, \\
    &         &   & mirror one end \\ \hline 
\becal~ Side/Top/Bottom &  3986     & 4288 & cut, ice polish \\ \hline
\becal~ Top/Bottom &  1583     & 6144 &  cut, ice polish, \\
    &         &   & mirror one end \\ \hline 
\dsecal         & 2144      & 2040 & cut, ice polish \\ \hline
 \pecal~ Side/Top/Bottom & 2410     & 1740 &  cut, ice polish,\\ 
    &         &   & mirror one end \\ \hline 
\end{tabular} 
\end{center}
\end{table} 

A procedure of automatic scanning of fibres was put in place primarily to
identify those fibres with poor light yield  that do not
necessarily show obvious signs of damage by a visual inspection. A secondary
function of the scanning was to measure and monitor the attenuation length of
the fibres which could be used as part of the \ecal~ calibration task. 
Two scanning methods were developed. The first, known as the Attenuation
Length Scanner (ALS), tested fibres one at a time and performed the QA
for the \dsecal~ consignment of fibres. The ALS was 
later replaced by the Fracture Checking Scanner (FCS) which had a faster
through-put of fibres as demanded by the construction schedule but with a less
precise measurement of the attenuation length. In addition, each \ecal~ module
construction centre ran a scan of all fibres within a module layer
as part of the construction procedure (as described in section \ref{sec:assembly}) in order to check that the
quality had not been compromised by the 
installation process.        

\begin{figure} 
	\centering
	\includegraphics[scale=0.4]{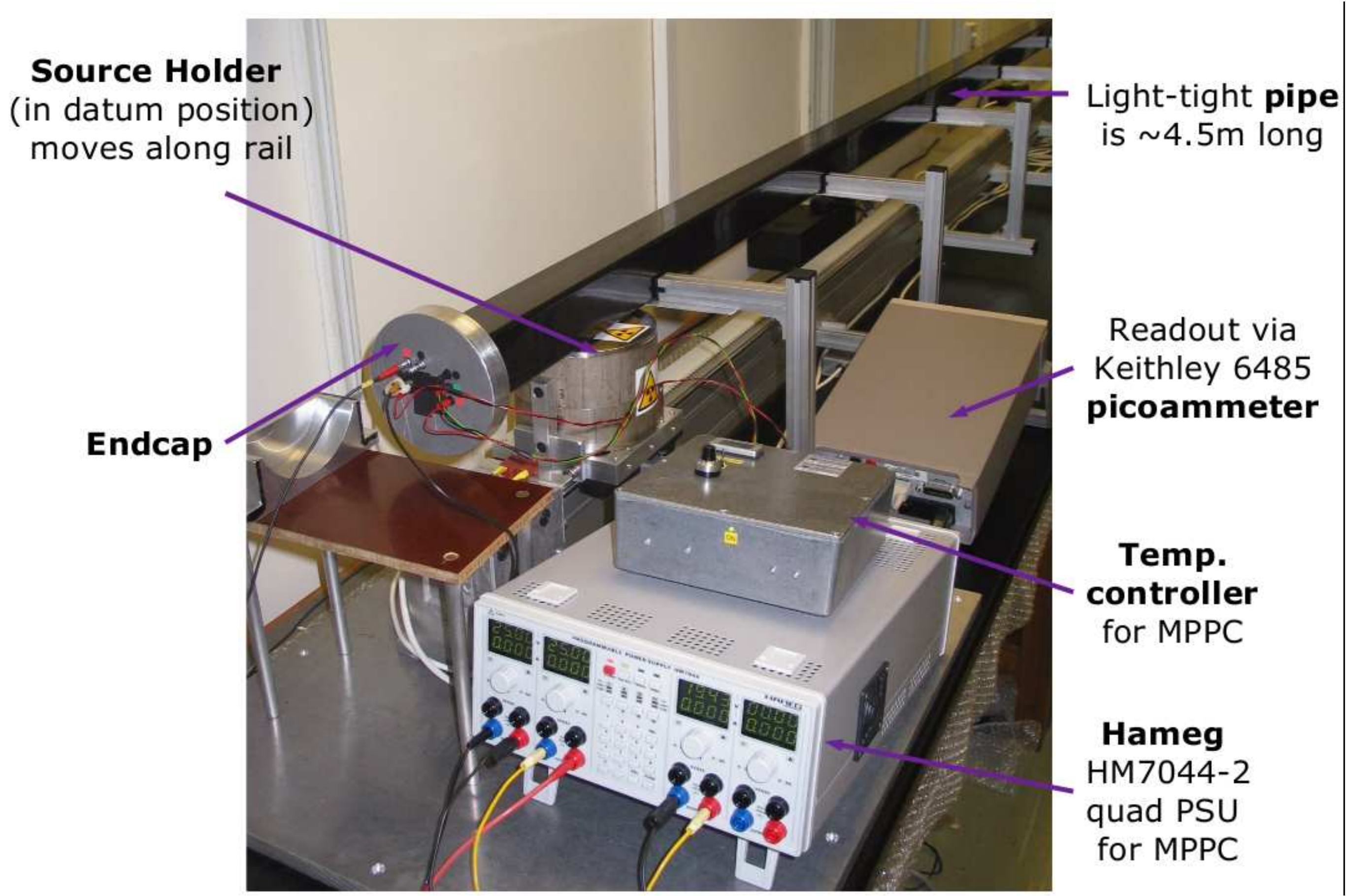} 
	\caption{\label{ALS}The ALS.}
\end{figure}

The ALS, shown in figure \ref{ALS}, consisted of a light-tight tube
containing a scintillator bar into which the test fibre is
inserted. A $5 \,$ mCi $^{137}$Cs source producing $662 \,$ keV photons was mounted
on a rail system underneath the tube and was able to travel over the
full length 
of the fibre under the control of a PC running LabVIEW.  The light output of
the fibre was recorded using a standard \ecal~ MPPC (as described in
section \ref{material-mppcs}) 
powered by a Hameg HM7044-2 quadruple PSU.   The output current was measured
by a Keithley 6485 pico-ammeter also under LabVIEW control.

\begin{figure} 
	\centering
	\includegraphics[scale=0.4]{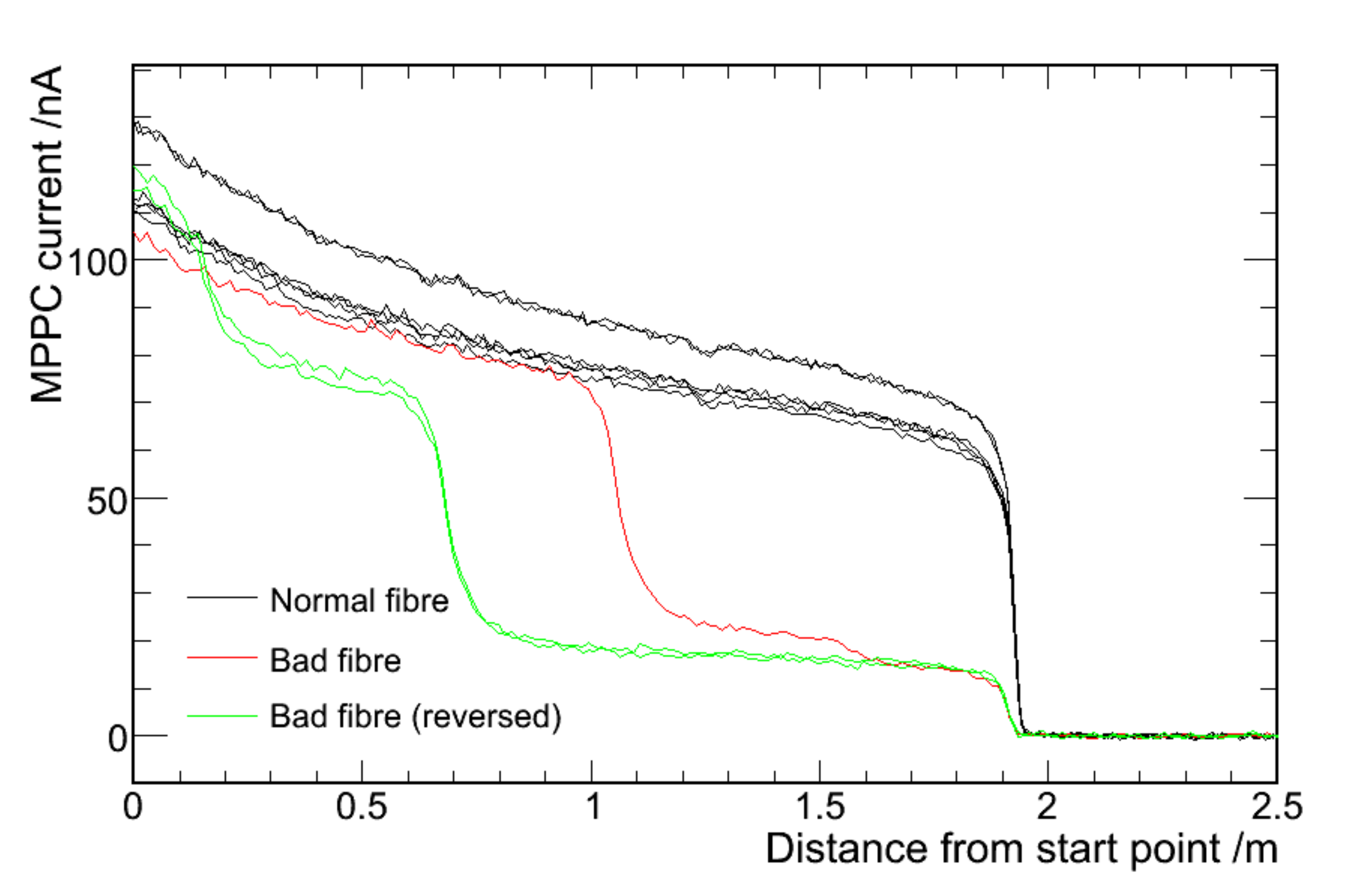} 
	\caption{\label{ALSresults}Some example ALS
          results of normal 
          fibres compared to a fibre exhibiting light loss.}
\end{figure}

 Figure
\ref{ALSresults} shows some typical results for good fibres compared to a bad
fibre  which shows light loss at two positions along its length (note that scans with
the fibre direction reversed help to confirm the locality of the light
loss). Such scans were performed on all of the fibres used for the
\dsecal~ and resulted in the rejection of approximately 100 out of a total 
of 2000  fibres.   

\begin{figure} 
	\centering
	\includegraphics[scale=0.4]{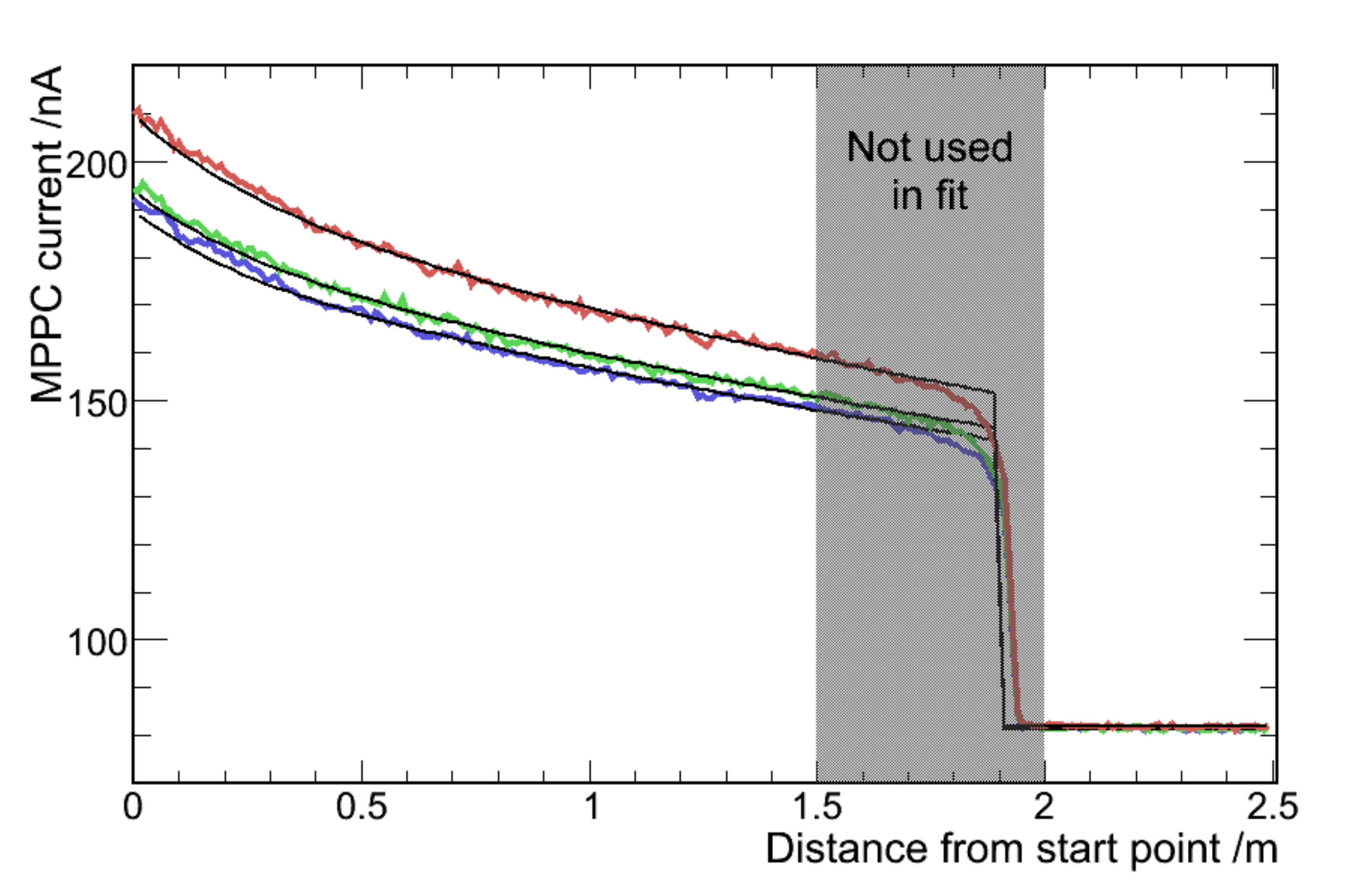} 
	\caption{\label{ALS2}Example fits to attenuation length data
          taken by the ALS. Coloured, jagged lines are data; solid,
          black lines are the fits.  The shaded grey region is not
          used in the fits.} 
\end{figure}

These data were also used in order to extract a measurement of the
attenuation lengths of the fibres.  
An example of attenuation measurements made with this apparatus can be seen in
figure~\ref{ALS2}. The current recorded from the MPPC in the
region up to $x=1.5$ m from the scan start point is
fitted to the following functional form \cite{WLSFormula}:
\begin{eqnarray}
I_{MPPC} & = & A \left( \frac{1}{(1+R)}e^{-x/\lambda_1} +
  \frac{R}{(1+R)}e^{-x/\lambda_2}   \right) + B.
\end{eqnarray}
The data from the \dsecal~ fibres suggested a flat background of
$B=81$ nA, $A=[108,128]$ nA, $R=[0.12,0.14]$ and attenuation length
coefficients of $\lambda_1=[3.9,4.1]$ m,  $\; \lambda_2=[0.21,0.31]$ m. 

\begin{figure} 
	\centering
	\includegraphics[scale=0.5]{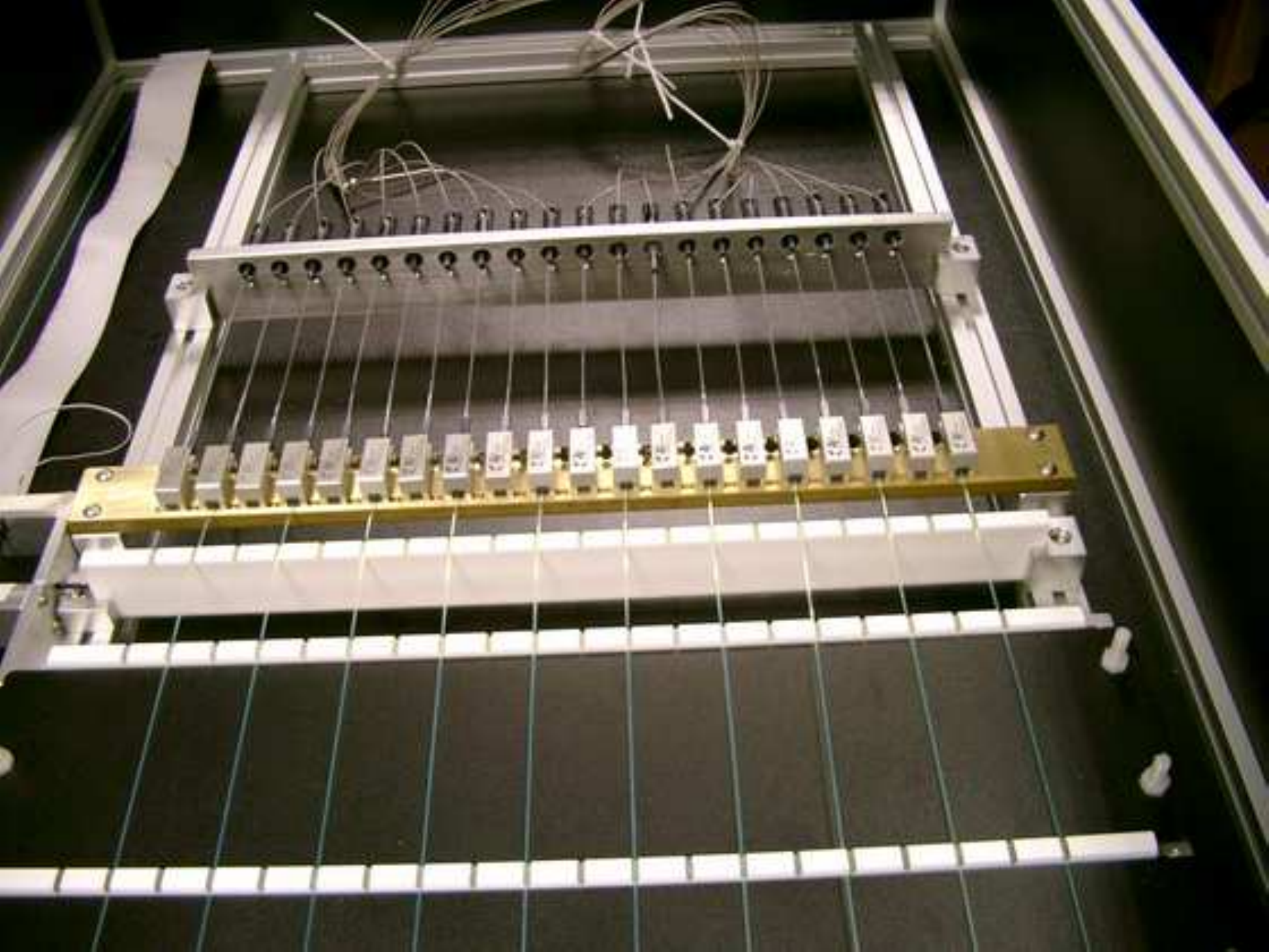} 
	\caption{\label{FCS}The FCS.  Ten
          wavelength shifting fibres can be seen lying in the grooved
          scintillator bars. The fibres connect to the MPPCs (shown in
          their black plastic housings) via a fibre clamp and a short
          section of clear optical fibre. The electronics are
          positioned behind the far end of the light-tight box.} 
\end{figure}

The FCS, shown in figure~\ref{FCS}, was
designed to scan twenty fibres quickly in a single run 
while still retaining a reliable identification of problem fibres. The scanner
was enclosed within a light-tight box with the readout electronics positioned
outside. A series of 29 scintillator bars were placed perpendicular to the
fibre direction and positioned over a distance of $4\,$m. The spacings of the
bars were selected to ensure that all of the four lengths of fibre had a
scintillator bar within $5\,$cm of the end of the fibre. Each scintillator bar
was cut with  20 v-shaped grooves into which the fibres were
laid allowing absorption of light from the bars.

Each scintillator bar was illuminated using an ultra-violet LED that was
coupled to a `leaky fibre',  one  with the cladding intentionally
  scratched to allow light to escape at points along the length, that was threaded through the bar. The UV light was absorbed by the scintillator bar which then emitted blue light out through the v-shaped grooves and into the WLS fibres. The UV LEDs were connected to a computer-controlled switching unit that used a binary addressing system to switch on each of the LEDs in turn.
The 20 WLS fibres were each coupled to \ecal~ MPPCs which were read out via a
multiplexer by the pico-ammeter.  The complete scanner system was controlled by
a single LabVIEW process.  

Software was used to check the WLS fibres for sudden drops in light
output between each of the illumination points, an indication that the
fibre was cracked or damaged in some way. One of the 20 WLS fibres in
each run was a reference fibre that had been scanned in the ALS. A
reference fibre was used for two reasons. Firstly, it allowed the
scanner to be calibrated such that the expected amount of light
collected at each position along the bar was known, allowing the WLS
fibres to be compared to the reference fibre. Secondly, it permitted
the attenuation length to be roughly measured by scaling the
attenuation profile of the reference fibre with the relative light
response from the other fibres.  The final yield from the QA steps of
WLS fibres delivered from FNAL, including any rejections due to the
ferrule gluing step described in section \ref{ferrules}, was 350
fibres rejected from the total order of approximately 17,000, a
rejection rate of about 2\%.

\subsection{Photosensors}\label{material-mppcs}

As mentioned in section \ref{design}, light produced in the
scintillator bars is transported via WLS fibres to solid state
photosensors.  Because the \ecal~ modules sit inside the iron return
yoke of the refurbished UA1 magnet, either the photosensors needed to
work inside the 0.2 T magnetic field provided by the magnet and be
small enough to fit inside the \ecal~ modules, or the light signal
would need to be transported several metres via optical fibres to
light sensors outside the ND280.  The use of MPPCs instead of
traditional PMTs allowed the first option to be chosen.  MPPCs consist
of many independent sensitive pixels, each of which operates as a
Geiger micro-counter.  The use of Geiger-mode avalanches gives them a
gain similar to that of a vacuum PMT. The output of the device is
simply the analogue sum of all the fired pixels, and is normally
expressed in terms of a multiple of the charge seen when a single
pixel fires, sometimes referred to as a `pixel energy unit', or PEU.  A
customized 667-pixel MPPC, with a sensitive area of $1.3 \times 1.3$
mm$^2$, was developed for T2K by Hamamatsu \cite{mppcCustomize}.  

In addition to meeting the above criteria, the MPPCs have a higher
photon detection efficiency (PDE) than PMTs for the wavelength
distribution produced by the WLS fibres.  Typical PDEs are given by
Hamamatsu for S10362-11-050C MPPCs~\cite{hamamatsuDataSheet}, which 
are similar to the customized MPPCs used by T2K. The peak efficiency,
which is at a wavelength of 
440~nm,  is around 50\%, and at the wavelengths emitted by the WLS
fibres (peaked around 510~nm), the efficiency is approximately
40\%. However, these PDE measurements were made using the total
photocurrent from the MPPC, and will therefore count pulses caused by
correlated noise such as crosstalk and afterpulsing in the same way as those
due to primary photons. More sophisticated analyses performed by T2K
in which these effects are removed give PDEs of 31\% at wavelengths of
440~nm, and
24\% in direct measurements of WLS fibre light~\cite{thesisMWard}.
Table \ref{tab:mppcs} shows the main parameters of the MPPCs.  More
information about the MPPCs can be found in references
\cite{t2kExperiment, fgdNIM, mppcPaper, mppcPaper2} and references
therein.

\begin{table}[h!]
\begin{center}
  \caption{\label{tab:mppcs}Main parameters of the T2K MPPCs. The dark
    noise rate is given for a threshold of 0.5~PEU, or half the charge of a
    single pixel firing.}
\begin{tabular}{|l|c|}
\hline
  Parameters & \\ \hline
  Number of pixels &667\\
  Active area  & $1.3 \times 1.3$ mm$^2$ \\ 
  Pixel size   & $50 \times 50 \; \mu$m$^2$ \\
  Operational voltage & $68 - 71$ V \\
  Gain   & $\approx 10^6$ \\
  Photon detection efficiency at 525 nm & 26-30\% \\
  Dark rate above 0.5 PEU, at 25$^{\circ}$C & $\leq 1.35$ MHz \\
  \hline
\end{tabular}
\end{center}
\end{table}

As T2K was the first large-scale project to adopt MPPC photosensors,
considerable effort was made to test the first batches of MPPCs before
detector assembly. Device properties were measured in a test stand
comprising 64 Y11 WLS fibres, illuminated at one end by a pulsed LED and
terminated at the other by ferrules connected to the MPPCs under test,
as described in section \ref{ferrules}. MPPCs were read out using a
single TFB board and a development version of the ND280 DAQ
software. Two such test stands were created, and the MPPC testing
(around 3,700 devices total) was divided equally between them.

The QA procedure consisted of taking many gated charge measurements
for each photosensor at a range of bias voltages, with and without an
LED pulse present during the charge integration gate. For each bias
voltage, a charge spectrum was produced from the measurements, and
analyzed in order to extract the sensor gain, similarly to the process
described in section \ref{calibration}.
The thermal noise rate, and
contributions from after-pulse and crosstalk, were extracted from the
relative peak heights in the charge spectrum, and a comparison of the
signals with LED on and off permitted the  extraction of the PDE;
absolute calibration of the incoming light level was  
performed using a MPPC whose PDE had been previously measured using
an optical power meter \cite{thesisMWard}.

The gain curve was fitted to calculate a pixel capacitance and
breakdown voltage for each device, and also the bias voltage required
to achieve a nominal gain of $7.5\times 10^5$. The PDE and noise
characteristics at this gain were then interpolated, to quantify the
performance of each device. All devices were found to be functional
and to perform acceptably (with reference to table \ref{tab:mppcs});
however, a 10\% contingency of sensors was ordered, and so devices were
rejected starting with those with the highest thermal noise rate. More
details on the QA procedure and its results can be found in
\cite{thesisHaigh}.  In situ, the dominant contribution to the
non-linearity of the combined scintillator/fibre/MPPC system is from
the MPPCs, and is estimated to be 2-3\% for MIPs and 10-15\% for
charge deposits typical of showers.

\subsection{Fibre to sensor coupling \label{ferrules}}

\begin{figure} 
	\centering
	\includegraphics[scale=0.4]{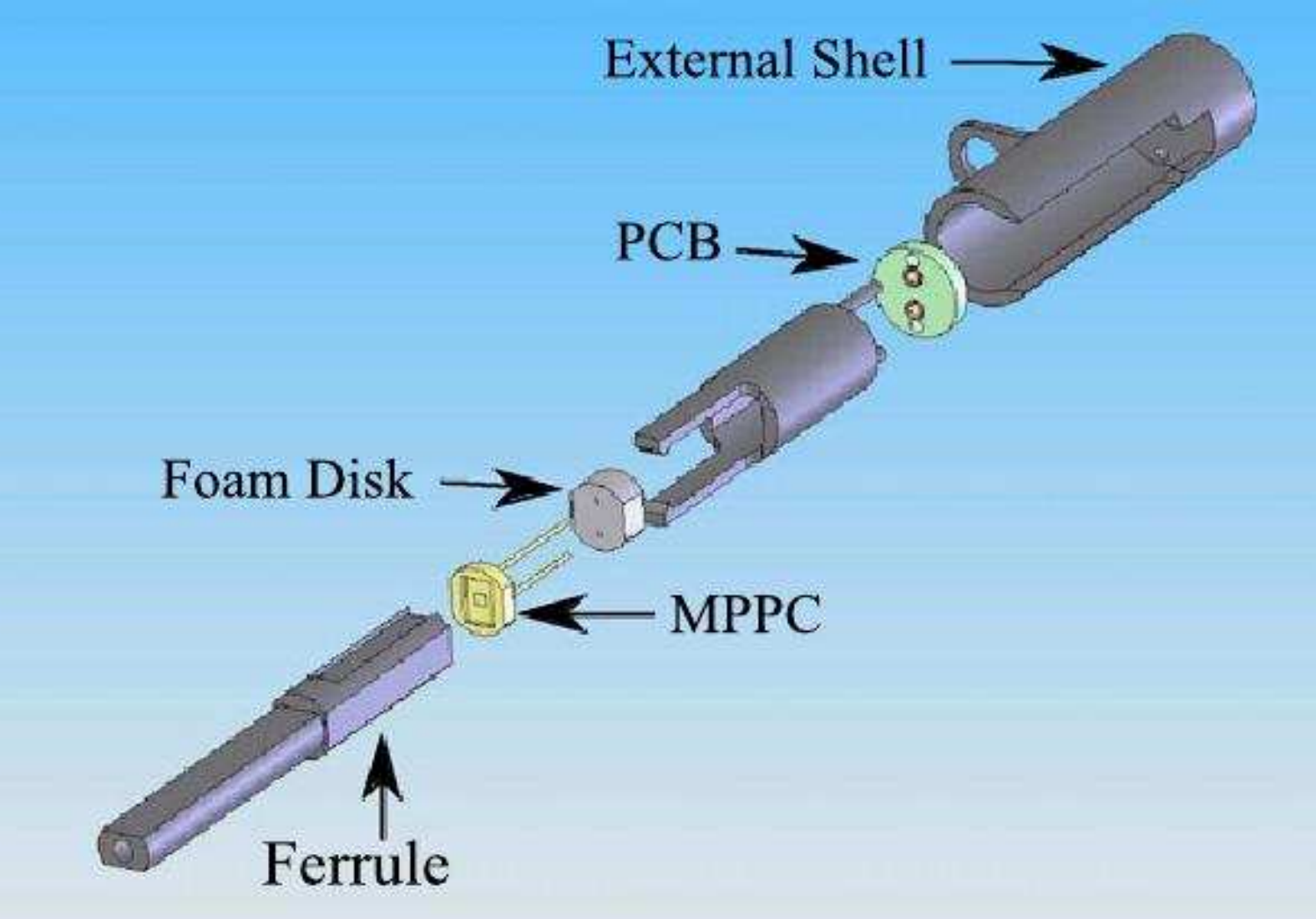} 
	\includegraphics[scale=0.9]{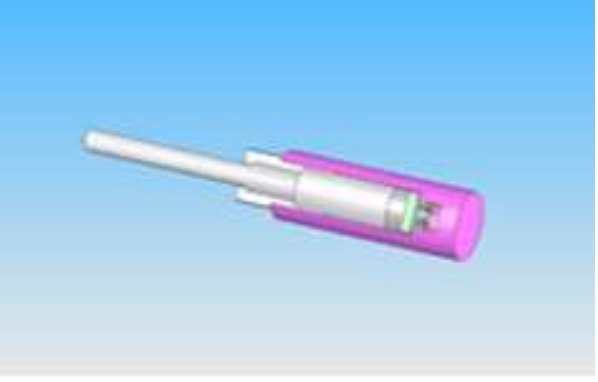}
	\caption{\label{Connector}(Left) An exploded view of the WLS fibre to
          MPPC coupling connector system. The ferrule in the bottom
          left of the figure is placed over the end of the WLS fibre
          (not shown), with the fibre overhanging the ferrule by
          0.5~mm in order 
          to ensure good coupling with the MPPC. The housing for the 
          MPPC and foam 
          spring is shown in the centre of the figure.  The `ears' of
          the housing slip over the ridge in the ferrule to lock the
          assembly together.  The external
          shell, or sheath, covers the MPPC housing and part of the
          ferrule when the connector is assembled, preventing the
          `ears' from disengaging with the ferrule.  The circular loop
          on the external shell allows the entire assembly to be screwed to
          the bulkhead securely.  (Right) The
          connector fully 
          assembled.  The assembled connector is approximately 5~cm long.}
\end{figure}

An essential component in the overall light-collection efficiency of the \ecal~
modules was the coupling of the WLS fibres to the MPPCs. The design used was
a multi-component solution shown in figure~\ref{Connector} (left) and was adopted
by the on-axis INGRID detector \cite{ingrid} and the \pod~ subdetector
\cite{pod} 
of the ND280 
in addition to the \ecal.  The 
assembly consists of three injection moulded\footnote{All injection
  moulded components were fabricated
  from Vectra \textregistered A130.} parts: (1) a `ferrule' glued to the
end of the fibre which engages with (2) a housing that holds the MPPC
and  ensures 
alignment with the fibre end to better than $150 \, \mu$m,  and (3) an external
shell, or sheath,  to contain the inner assembly and provide protection. A $3 \,$mm-thick
polyethylene  foam disk  
sitting just behind the MPPC provided sufficient contact pressure
between the fibre end and the MPPC epoxy window to ensure an efficient
connection 
without the use of optical coupling gels which could deteriorate over time and
present a complicated calibration challenge.  
Electrical connection between the MPPC and the front-end electronics is provided  
by a small, circular, printed circuit board with spring-loaded pin sockets 
which 
contact the legs of the MPPC and connect to a micro-coaxial connector by Hirose
(not shown in
figure \ref{Connector}). Figure~\ref{Connector} (right) shows the connector fully
assembled. Early prototypes of the connector revealed that an
unacceptable light 
loss could occur if the fibre end was glued slightly short of the ferrule
end. This led to the production of gluing guides which ensured a precise
overhang of the fibre from the ferrule end by $0.5 \,$mm with a high production
reliability. 

\section{Construction \label{construction}}
The T2K UK group employed a distributed construction model to optimize
efficiency, space and the use of available personnel.  Module layers
were constructed first and then lowered one at a time inside prepared
bulkheads, after which the MPPCs, then cooling panels, TFBs, cooling
pipes, and finally cover panels were attached.  The RMMs were then
affixed to the outside of the modules.  As each layer was installed
inside the bulkheads, a two-dimensional (2D) scanner, discussed in
section \ref{scanner}, carrying a 3~mCi~ $^{137}$Cs source was used in
conjunction with a well-understood set of `test' MPPCs to check the
integrity of the bar-fibre combination before the next layer was
installed, enabling repairs to be made if necessary.  The material
preparation and QA also followed a distributed pattern.  The
distribution model is shown in table \ref{tab-constructionModel}.  All
of the components in the table had to come together on a co-ordinated
schedule for the modules to be constructed.  Details of the layer and
module assembly are given in the following sections.

\begin{table}
\begin{center}
\small
\caption{\label{tab-constructionModel}The \ecal~ construction and QA
  model, showing contributions from Daresbury Laboratory (DL), Imperial
  College London, Lancaster University, Liverpool University,
  Queen Mary University 
  London (QMUL), Rutherford Appleton Laboratory (RAL), University of
  Sheffield, and University of Warwick.}
\tabcolsep=0.11cm
\begin{tabular}{|l|c|c|c|c|c|c|c|c|} \hline
     &DL &Imperial &Lancaster &Liverpool &QMUL &RAL &Sheffield &Warwick \\
\hline
\ecal~  design                & X &   &   & X &   &   &   & \\
Module engineering/constr.    & X &   &   & X &   &   &   & \\
\dsecal~ layers               &   &   & X &   &   &   &   & \\
\dsecal~ module              &   &   & X &   &   &   &   & \\
\becal~ side layers           & X &   &   &   &   &   &   & \\
\becal~ side modules          & X &   &   &   &   &   &   & \\
\becal~ top/bott layers       &   &   & X &   &   &   &   & \\
\becal~ top/bott modules      & X &   &   & X &   &   &   & \\
\pecal~ side layers           &   &   &   &   &   &   &   & X \\
\pecal~ side modules          &   &   &   &   &   &   &   & X \\
\pecal~ top/bott layers       &   &   &   &   &   &   & X & \\
\pecal~ top/bott modules      &   &   &   &   &   &   &   & X \\
2D scanner                    &   &   &   &   & X &   &   & \\
MPPC QA                       &   & X &   &   & X &   &   & X \\
Scintillator bar QA           &   &   & X &   &   &   & X & \\
WLS fibre QA                  &   &   &   &   &   &   &   & X \\
MPPC-WLS connectors           &   &   &   &   &   &   &   & X \\
Electronics                   &   & X &   &   &   & X &   & \\ \hline

\end{tabular}
\end{center}
\end{table}

\subsection{Layer assembly}

Each layer is framed by aluminium bars with an L-shaped cross-section
of dimensions 20.00 mm (base) $\times$ 12.54 mm (height).  The height
of the base is approximately 10.2 mm, very slightly higher than the
scintillator bars, and the width of the stem is
20.00 mm.  Construction of the layer began by screwing the aluminium
bars into place onto a Teflon-covered assembly table.  The
scintillator bars then were prepared by applying a two-part epoxy
(Araldite 2011 Resin and Araldite 2011 Hardener) to one edge of the
bars.  The bars then were laid inside the layer frame such that the central
hole of each bar was aligned with a 2 mm-diameter hole in the frame.
An O-ring with an 
uncompressed thickness of 1.5 mm was inserted into the 1 mm gap
between the ends of the bar and the layer frame and compressed into
place.  This was done to prevent epoxy from entering the bar hole and
compromising the 
subsequent insertion of WLS fibre.  The position of each bar was
stabilized during layer construction by inserting a temporary tapered
Teflon-coated locator pin through the frame hole, the O-ring, and into
the bar hole. The locator pins were removed when the layer was
complete.

\begin{figure}[h]
\begin{center}
\includegraphics[width=\textwidth]{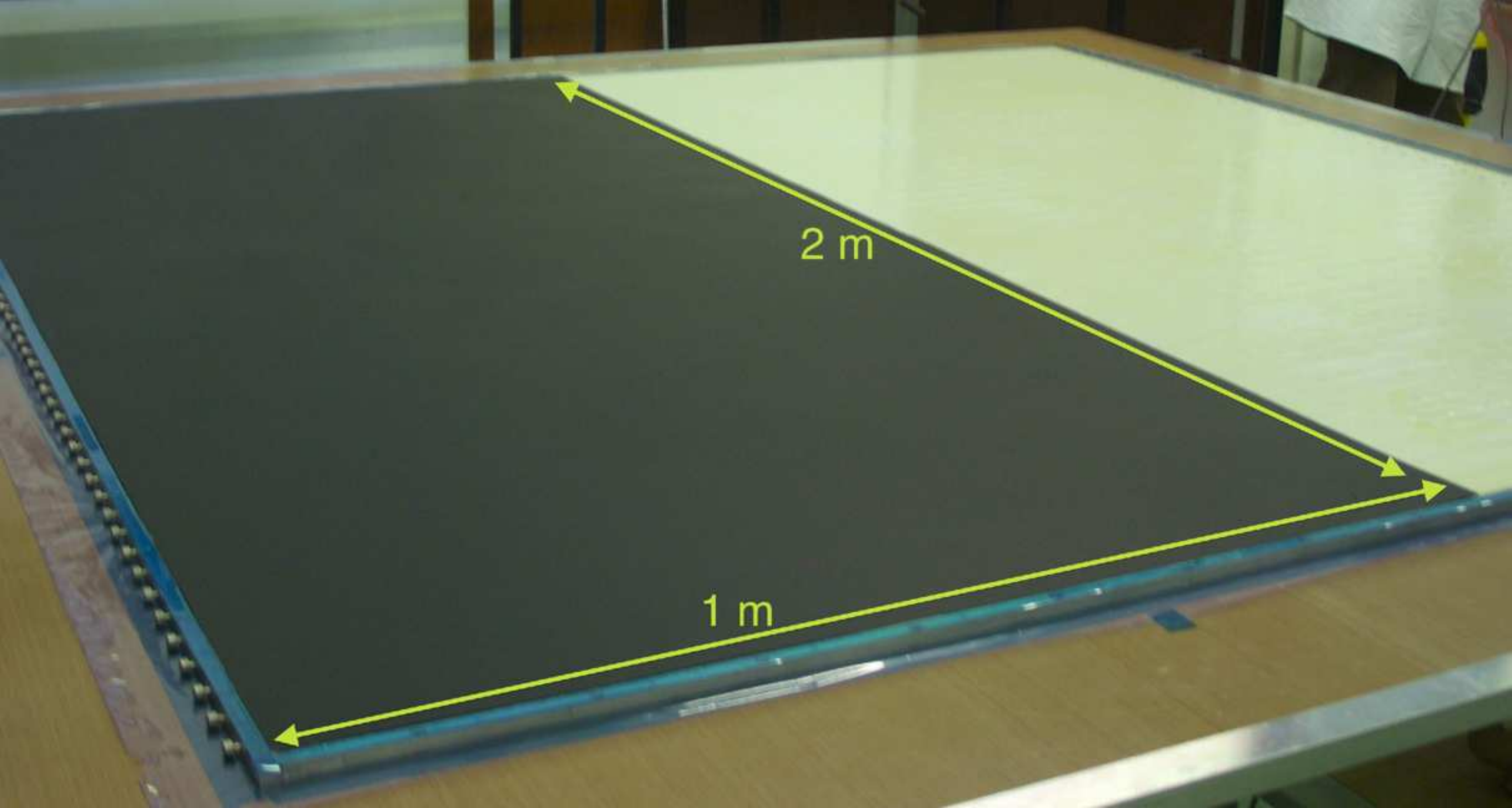}
\caption{\label{fig-layerPartDone}
\dsecal~ layer under construction.  The first of two sheets of lead is
in place on top of the scintillator bars.  Visible are the aluminium
frame and the locator pins securing the scintillator bars in place for
the duration of the layer construction.  The frame is covered with
blue tape to keep it free from epoxy.}
\end{center}
\end{figure}

Once all of the bars were stabilized in position, a thin layer of
epoxy was applied to them and to the lip of the frame, and the lead
sheets were placed on top, using a vacuum lifting rig attached to an
overhead crane in order to  distribute the
weight across several equally-spaced suction cups and so 
avoid distorting the lead. The sheets were carefully positioned to
minimize the gap between them, thereby avoiding a region
of low 
density in the middle of the modules, while ensuring sufficient overlap
onto the lip of the layer frame to maintain structural integrity.
Figure \ref{fig-layerPartDone} shows one of the \dsecal~ layers 
being constructed, with one sheet of lead in place on top of the
scintillator bars.  

The entire layer then was covered with a sandwich of vacuum-sealing
plastic and fabric, with the top layer of plastic 
securely taped to the table.  Vacuum pumps  were used to evacuate the
air around the layer,
allowing the epoxy to cure for 12 hours under vacuum compression.  

Once the curing was finished, the layer was unwrapped, the locator
pins removed, the screws securing the layer frame to the table were
removed, the WLS-fibre holes were tested to ensure that they had not been
blocked with epoxy, and the layer then was stored for use in a module.

\subsection{Assembly procedures for the \ecal~ modules \label{sec:assembly}}

The \dsecal~ was the first module to be constructed and most of the
procedures developed during the process were used on the other modules
as well.  The first step was to assemble the bulkheads and the
carbon-fibre panels.  One carbon-fibre panel (the bottom panel during
construction which would become the upstream face when the \dsecal~
was in situ) was attached to the bulkheads to form an open box.  The
other (top) carbon-fibre panel was stored until later.  The bulkhead
box was positioned on the construction table, and the 2D scanner,
discussed in section \ref{scanner}, was attached and commissioned.
The first layer then was lowered inside the bulkheads and positioned
on top of the carbon-fibre base.  A 1 cm gap between the bulkheads and
the layer on all four sides was obtained by tightening or loosening
grub screws which were inserted through holes in the bulkhead and
tensioned against the layer frame.  The LI LED strips and perspex 
lenses (see section \ref{li}) then were glued onto the bottom
carbon-fibre panel in the 1 cm gap; the LI electronic cards were affixed to
the inside of the bulkheads, with the LI cables routed outside the
bulkheads through the air holes.  WLS fibres were inserted through the
scintillator bars.  A MPPC-fibre connection ferrule was bonded to each
fibre using Saint-Gobain BC600 silicon-based optical epoxy resin.  The
test MPPCs were coupled to the fibres using the connection sheaths and
connected via a mini-coaxial cable to TFBs, which provided the control
and readout (see section \ref{electronics} for a description of the
TFBs). After this the layer was covered and made light-tight, and a 2D
scan was taken.

\begin{figure}[h]
\begin{center}
\includegraphics[scale=0.65]{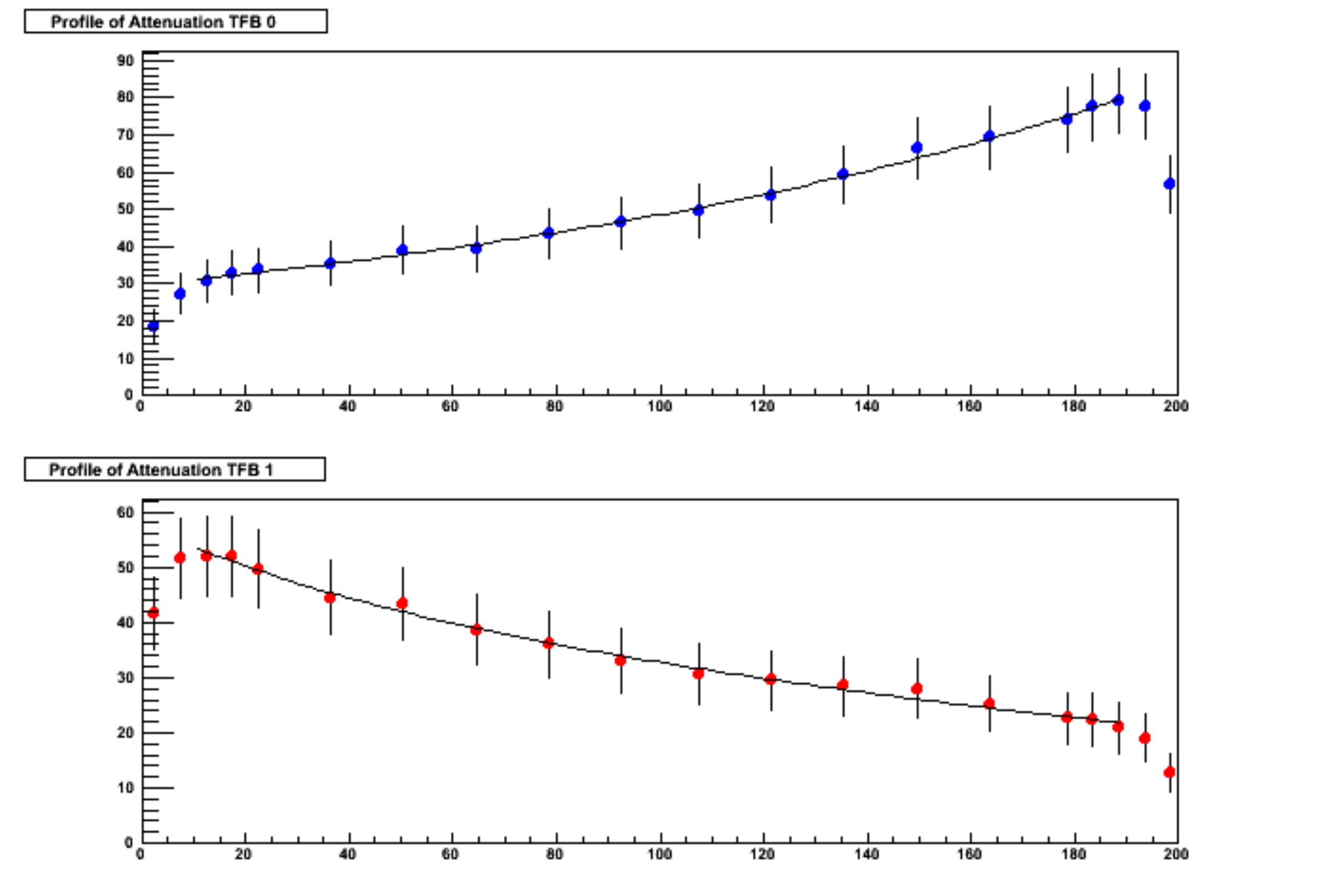}
\caption{\label{fig-scannerPlot}
A typical light attenuation profile from scanner data corresponding to
one scintillator
bar in the \dsecal.  The ordinate axis is the ratio of the integrated
light yield with the source present to the integrated light yield
without the source; the
abscissa is the position along the bar in cm.  The light yield
measured by the MPPC at one end 
of the bar and read out by one of the two TFBs (TFB 0) is shown in the
upper plot,  and that from the MPPC at the
other end read out by the other TFB (TFB 1) is shown in the lower
plot.  The points are data; the curves 
are a single-exponential fit to the central region of the data.}
\end{center}
\end{figure}

The 2D scanner collected data at 20 points along each 2000 mm bar,
with data points being closer together near the ends in order to
facilitate 
an understanding of the light escaping through the ends of the
scintillator bar.  For efficiency, the analyzing software ran in
parallel with the data-taking, producing an attenuation profile for each
bar in the layer.  A typical example of this is shown in figure
\ref{fig-scannerPlot}.  The ordinate axis shows a reference value of
the light yield since it is calculated as a ratio of the integrals
(from 5.5 PE to 30 PE) of the MPPC
response when the source is present, to the response when the source
is not present, and therefore represents (signal +
background)/background.  This ratio is calculated at each data point
along the 
length of the bar.  More information about the analysis of the scanner
data is available in \cite{thesisDavies}.

The scan data were checked and if problems were encountered,
appropriate action was taken.  A common problem involved the coupling
between the fibre and the MPPC, often due to the difficulty of
positioning the ferrule on the fibre.  In this case, since the ferrule
could not be removed, the fibre would be replaced, a new ferrule would
be attached, and the bar would be re-scanned. After this process, the
test MPPCs were removed and the next layer was installed and scanned
in the same manner.  Where required, thin Rohacell foam sheets were
placed between the layers to ensure that the layers did not warp
inwards.

After several layers were installed in the \dsecal, it was noted that
the holes in the bulkheads were no longer aligning well with the holes
in the layers and the scintillator bars, which made the insertion of
the WLS fibre difficult.  Measurements indicated that the layer frames
had been incorrectly manufactured and were an average of 0.2 mm higher
than the specifications.  This was remedied by using a router to thin the frames on
the layers that were not yet installed.  The layer frames for
subsequent modules did not have this problem.  

After all of the layers were installed, LI LEDs and electronic cards
were attached to the top carbon-fibre panel.  9 mm of Rohacell foam
was glued to the inside of the panel to ensure that the layers within
the bulkheads stayed stable when the \dsecal~ was in its upright
position in situ.  The carbon-fibre panel was then affixed to the top
of the bulkheads.

The procedure described above completed the construction of the active
region of the detector.  The next steps dealt with the data readout.
First, the 3400 \dsecal~ `production' MPPCs in their custom-made
sheaths complete with foam springs and mini-coaxial cables were
attached to the ferrules of every layer and secured to the bulkheads
as described in sections \ref{dsecal} and \ref{ferrules}.
Cable-management brackets were attached to the bulkheads and the MPPC
cables were grouped together and tidied in preparation for the next
steps.  Tyco Electronics LM92 temperature boards were screwed onto the
bulkheads between the MPPCs in positions that allowed one temperature
board to be connected to each TFB.

The cooling panel for the left-side \dsecal~ was assembled from four
separate cooling plates.  The TFBs then were attached and thermally
connected to the left-side cooling panel using screws and
thermally-conducting epoxy resin.  The cooling panel then was held in
position on the left side of the \dsecal~ while the MPPC and
temperature-board cables were threaded through the slots in the panel
and the LI cables were routed through the panel's air holes, after
which the panel was placed into its final position and bolted into
place.  The MPPC and temperature-board cables were connected to the
correct ports on the TFBs.  The same procedure was followed for the
other three cooling panels.

The water-cooling circuit then was installed on the cooling panels and
the gas distribution branch was installed on the bottom cooling panel.
These systems were tested under pressure.  Following this,
cable-management brackets were fitted between all of the TFBs, and
low-voltage feedthroughs and bus bars were installed.  The bus bars
were checked for continuity, isolation from ground and from each
other.  Shielded Cat~5e cables then were connected to the
TFBs. Along with the LI cables, they were routed around the detector
to a patch panel mounted on the bottom cooling panel or, in the case
of the trigger cables from the TFBs and half of the LI cables, to the
top cooling panel.  Figure \ref{fig-dsside} shows the left-side
\dsecal~ cooling panel with the TFBs, bus bars, Cat~5e cables, LI
cables, water-cooling pipes and air holes.

The outer cover panels then were attached to each side.  The top cover
panel was fitted with 28 cable glands through which the Cat~5e cables
corresponding to the trigger system exited the detector, and with air
vents to allow the gas being flushed through the detector to
escape. Half of the cables from the LI system exited through holes in
the top cover panel.  The bottom cover panel was fitted with power
cable clamps, cable-management brackets and RMM cards.  The Cat~5e
cables were routed from the patch panel to the RMMs.  LI junction
boxes were attached and connected to the LI cables, completing the
construction of the \dsecal.

This construction procedure was repeated for each of the \becal~
modules. Minor alterations to the method were needed to accommodate
the slightly different structure of the modules (see section
\ref{becal}).  More significant alterations were made to the method
for the \pecal~ modules, reflecting their slightly different design, as
described in section \ref{pecal}. The most significant differences
were that the TFBs were mounted directly on  the bulkhead instead of onto cooling panels, with the MPPC cables
  routed to them and supported 
  where necessary; the smaller number of TFBs per module allowed for
  the use of a thick standard copper wire, instead of bus bars, to supply
  power, which was distributed to the TFBs via branching connections
  to terminal blocks; in the final step of the assembly, a
  non-magnetic support structure was bolted to the aluminium back panel, which
  positions the thinner \pecal~ modules away from the magnet, and
  closer to the basket.

\subsection{The bar scanner \label{scanner}}

Three-axis scanners were designed to position a $^{137}$Cs radioactive source
at multiple points above the surface of each detector layer as each layer was
assembled into the subdetector body. Three variants of the scanner were
manufactured:
\begin{itemize}
 \item \dsecal~ and \pecal~ scanner, with a footprint of 3928~mm $\times$ 3578~mm; 
 \item side \becal~ scanner, with a footprint of 4862~mm $\times$ 3240~mm;
 \item top--bottom \becal~ scanner, with a footprint of 4862~mm $\times$ 2410~mm.
\end{itemize}

The three axes were driven by Mclennan SM9828 Stepper Motors
controlled by PM600 Intelligent Stepper Motor Controller.  
The controller was programmed via a commodity PC running LabVIEW. The PC and
stepper motors, along with their controller, were all powered via an
uninterruptible power supply (UPS). The system design included an APC
Smart-UPS 2200, 230V, primarily as a safety feature so the radioactive
source could be automatically parked during 
a power cut; however, it had the added benefit that short duration power
glitches did not stop a running scan.

These controls were integrated into a LabVIEW program providing an
operator interface to control the machine. This operator interface was
implemented as a `state' machine, the main states being `source loaded' and
 `source unloaded'. In the source-unloaded state the
radioactive source was not attached to the scanner head. 
In this state the
head was raised and moved to one side to position the arm in the least
inconvenient position for those working on the detector.
The arm could also be moved in the $x$-direction to allow for greater
access to the 
detector during construction.

The source was loaded under computer control with the computer prompting the
operator to perform the necessary steps in a safe order. The program then
switched to the source-loaded state where scanning parameters could be input
and the scan started. Each scan started with the $z$-arm searching for the
surface of the module layer at the centre of the module. The arm descended
slowly until the push rod mounted adjacent to the source operated. The arm
then backed off and the source moved to the matrix of measurement positions.
At each position the Scan Control program prompted the DAQ to start
data-taking via a network link. On completion of data-taking, the scanner would 
move to the next position and iterate until data had been taken at all
positions. When the data-taking was complete, or if the program detected an error
condition, the source would be returned to its lead-shielded safe parked
position. Unloading the source was again under computer control with operator
prompts to ensure safety. Re-positioning of the scanner head was found
to be accurate to within 0.02~mm.

\begin{figure}[h]
\begin{center}
\includegraphics[scale=0.50]{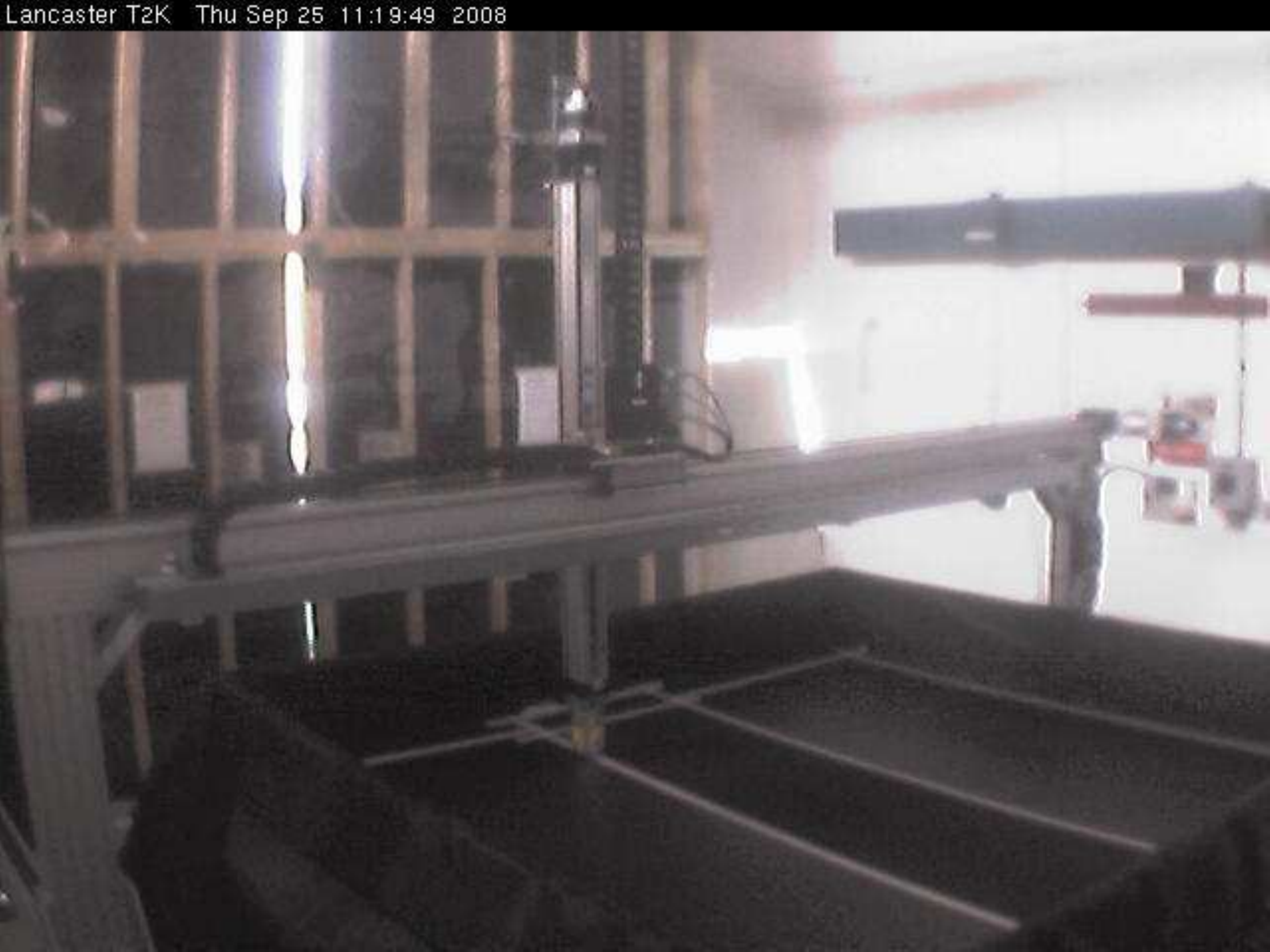}
\caption{\label{fig-scannerInMiddle}
The 2D scanner moving along the scintillator bars in the \dsecal.  The
$^{137}$Cs source sits at the bottom of the vertical arm of the
scanner, just above the layer being scanned.  The blackout material
covering the layer is seen, along with the tape holding it in place
for the duration of the scan.}
\end{center}
\end{figure}

Figure \ref{fig-scannerInMiddle} shows the scanner in operation during
the \dsecal~ construction.   The vertical arm finds its position as
described in the text.  The operator provides the $x-y$ coordinates
and the required timing at each position.  Since radiation safety rules dictated that
no one should be in the area during the scan, the image was taken by a
web camera which allowed operators to check on the scanner progress.

\section{Readout electronics and data acquisition}\label{electronics}
As described in section \ref{dsecal}, the mini-coaxial cable from each
MPPC is routed outside the cooling panels and connected to a custom
designed TFB.  Each TFB contains 4 Trip-T application-specific
integrated circuits (ASICs), originally designed for the D0 experiment
at FNAL. Up to 16 MPPCs can be connected to each ASIC, implying that a
maximum of 64 MPPCs can be connected to a single TFB.  In total, the
\ecal~ has 22336 electronic channels connected to MPPCs.  To increase the
dynamic range of the electronics, the incoming MPPC signal is
capacitively split (1:10) into high- and low- gain channels, which are
read by different channels of the ASIC. Depending on the MPPC gain,
the single-pixel (1 PEU) signal corresponds to approximately 10 ADC
counts in the high-gain channel, while the maximum signal in the
low-gain channel corresponds to around 500 PEU.

The Trip-T chip integrates the charge in a preset (programmable) time
interval which is followed by a programmable reset time at least 50 ns
long.  For T2K, the integration windows are programmed to synchronize
with the neutrino beam timing.  The Trip-T chip integrates from 23
readout cycles in a capacitor array and once all cycles have been
completed the stored data are multiplexed onto two dual-channel
10-bit ADCs which digitize the data.  Signals from the high-gain
channel are routed via a discriminator which forms part of the Trip-T
chip.  A field-programmable gate array (FPGA) produces timestamp
information from the discriminator outputs and sends this information
together with the ADC data to a back-end board.  In addition, the TFB
also records monitoring data (e.g. temperature, voltage) via the same
FPGA, which is asynchronously transmitted to the back-end board for
data concentration and buffering.  Detailed information about the
Trip-T chip and front-end electronics is given in references
\cite{t2kExperiment} and \cite{Vacheret2007}.

The back-end electronics system for the ND280 
consists of several different boards. The TFBs are connected to
RMMs which provide control and readout.
Control is fanned out from a master clock module (MCM), via several
slave clock modules (SCMs), one per subdetector.  Additionally, two
cosmic trigger modules (CTMs) are used to provide a selection of
cosmic-ray muon triggered events for calibration and monitoring.  All
of these boards use a common hardware platform, specifically developed
by the Rutherford Appleton Laboratory for use in the T2K experiment.
Signals from up to 48 TFBs, which are mounted on the detector
typically less than 1~m away from the MPPCs, are routed to one RMM via
Cat~5e cables.  The \ecal~ uses a total of 12 RMMs: 8 for the \becal,
2 for the \dsecal~ and 2 for the \pecal.  Each RMM controls its
associated TFBs, distributes the clock and trigger signals to them and
receives data from them once a trigger has been issued.  Data from the
RMMs are then sent asynchronously via a gigabit ethernet link to
commercial PCs that collect and process the data.

The ND280 uses a single MCM.  This receives 
signals from the accelerator which allow it to determine 
when a neutrino spill is about to occur, and also from a GPS-based clock 
which is used to synchronize the electronics to UTC.  The MCM 
prioritizes and issues triggers across the whole detector, and manages 
readout-busy situations. The signals and control of the MCM are fanned out 
to the SCMs. The trigger and clock signals are
passed to the \ecal~ RMMs via the \ecal~ SCM.  The SCMs allow the electronics 
for a sub-system to be configured independently.   It is possible for the 
\ecal~ to run autonomously (`partitioned DAQ') from the rest of the 
ND280 for calibration and debugging by using 
the \ecal~ SCM as master controller.  

The software control of the detector is performed using the
``Maximum Integration Data Acquisition System'' (MIDAS) \cite{MIDAS}.
The front-end software is custom written to manage the communication
with the RMMs, CTMs, SCMs, and MCM through gigabit ethernet links.  A
second process combines ADC and TDC information, compresses the data
and makes histograms for pedestal determination and monitoring.  A
third process manages communications with computers running the MIDAS
DAQ elements.  These three processes co-operate on front end PCs
running Scientific Linux, with each front end PC being connected to
two back-end boards.  The event builder and data logging use software from the
MIDAS distribution with virtually no customization for T2K.  The DAQ
contains an online monitoring system which makes histograms for
assessing data quality in real time and passes events to the online
event display for monitoring.  Detailed information about the DAQ is
given in \cite{t2kExperiment}.

\section{Light injection system}
\label{li}
The ND280 \ecal~ LI system is designed to provide a
quick and reliable method of
monitoring the performance of the MPPCs used inside the
\ecal~ modules. A complete discussion of the LI system R\&D can be found
in reference \cite{thesisGWard}. 
The LI system is required to illuminate the MPPCs on a given readout
face of an \ecal~ module, with a short duration optical pulse. The pulse
length and stability must be sufficient to afford accurate ($\approx$1 ns)
timing calibration. The intensity across the readout-face should be
uniform and any electromagnetic or electro-optical noise induced by
the system must not interfere with the surrounding sensors or
electronics.

In order to accomplish these aims, the LI system employs a modular design
incorporating dedicated electronics for both the interface with the
ND280 DAQ and the pulsing of LI sources within an \ecal~ module. 
The components are described in the subsections that
follow. The LI front-end electronics are housed in custom-built crate
assemblies compatible with a standard 19-inch rack. In brief the LI
signal chain comprises the following components:
\begin{enumerate}
\item The trigger card receives the ND280 MCM signal.
\item Control cards (CCs) receive and interpret the DAQ instructions and
  collate with the clock.
\item The junction boxes (JBs) receive and fan out the CC outputs.
\item The pulser receives the CC output via a
  JB and drives LED strips which emit optical pulses for
  calibration.
\end{enumerate}
Each of the components is described in more detail below.
\subsection{Control cards and trigger receiver}
\label{li:cc}
The LI system receives DAQ instructions into a dedicated CC. Each card
hosts a TCP/IP server that allows the DAQ 
instructions to be interpreted and then encoded into a
sequence of TTL pulses used to drive LED pulser cards, housed inside
the \ecal~ modules. 
The ND280 MCM transmits a $100$\,MHz signal to the LI system which is
received by a dedicated trigger card. The clock signal is
first fanned out to each CC, then collated
with the CC outputs onto a RJ45/Cat 5e signal cable ready for
distribution to the LI JBs.
\subsection{Junction boxes}
\label{li:jb}
The JBs are responsible for directing the TTL pulse
train and MCM 
clock from the CCs to the relevant LI pulsers, the pulser being the
dedicated electronics required to form the electrical excitation
pulses used to drive the LED strips used for illumination.
There is one JB per \ecal~ module except for the \dsecal~ which has two: $6\times$\,\pecal,
$6\times$\,\becal~  and
$2\times$\,\dsecal,
making 14 in total. The JBs are mounted outside, but in close
proximity to, the \ecal~ modules themselves. They are passive devices,
used exclusively for fanning out the CC pulses and therefore introducing
no modification to the actual signal.
\subsection{Communications protocol and cabling}
\label{li:comms}
The low-voltage differential signalling (LVDS) protocol is used
throughout to ensure 
robustness against interference from electromagnetic noise in
the detector. The DAQ instruction and MCM clock information are received over
LVDS, converted to TTL for interpretation by the CCs. The CC outputs are again
transmitted using LVDS, converted to TTL for fan out in the JBs and
then transmitted to the LI pulsers again using LVDS. 
The \dsecal~ portion of the LI system, however, is an exception in that it
features an older design which implements TTL over LEMO cables rather than
LVDS over RJ45/Cat 5e cables.
In order to preserve the relative timing between pulsers, all Cat~5e/LEMO
cables within an \ecal~ module have been installed with the same length.

\subsection{Pulsers}
\label{li:pulsers}
\begin{figure}[htbp]
  \centering
  \includegraphics[width=0.9\textwidth]{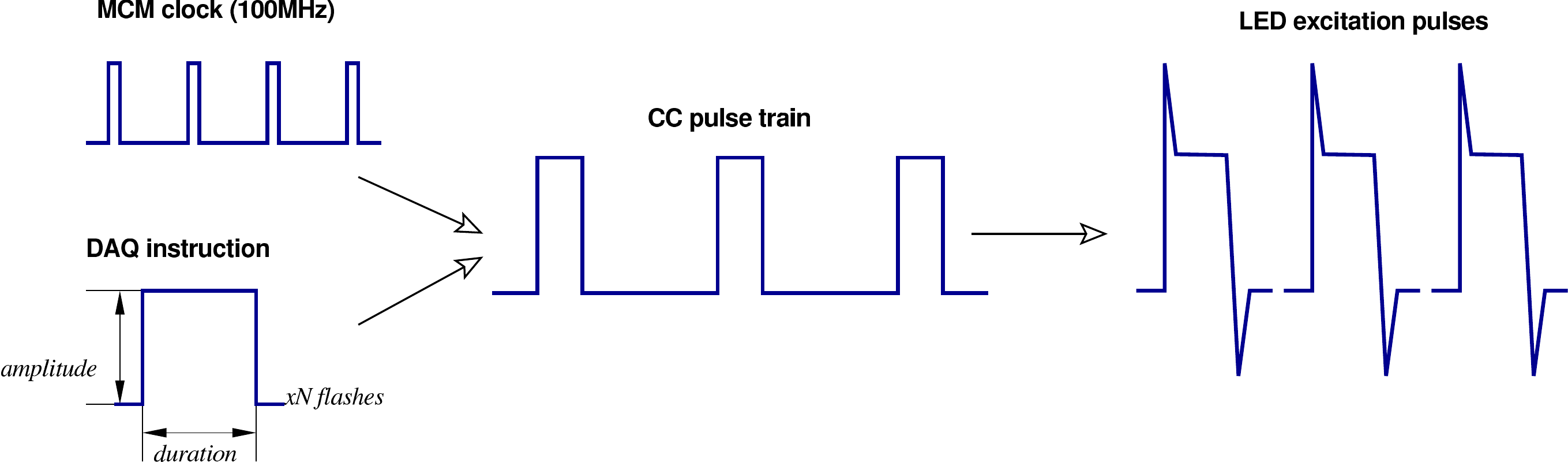}
  \caption{The LI signal path. The CCs encode an MCM
    synchronized pulse train with the desired timing and amplitude
    characteristics. The pulsers derive the excitation charges from
    the CC output.}
  \label{fig:signal}
\end{figure}

LED pulser cards mounted inside the \ecal~ modules receive the
pulses emitted from the CCs via the JBs. The TTL logic pulses
determine the pulse duration, amplitude and number of flashes. A
shaping component of the pulser board introduces activation and deactivation
spikes to the leading and trailing edges of a square-wave electrical
pulse. This ensures a constant level of illumination with a negligible optical
rise time.
The signal chain is illustrated in figure \ref{fig:signal}.
Different numbers of pulsers are located in different \ecal~ modules,
depending on their size and layout. There are 114 in total: 14
\pecal, 84 \becal~ and 16 \dsecal.
\subsection{LED strips and extruded perspex lens}
\label{li:leds}
\begin{figure}[htbp]
  \centering
  \includegraphics[width=0.39\textwidth]{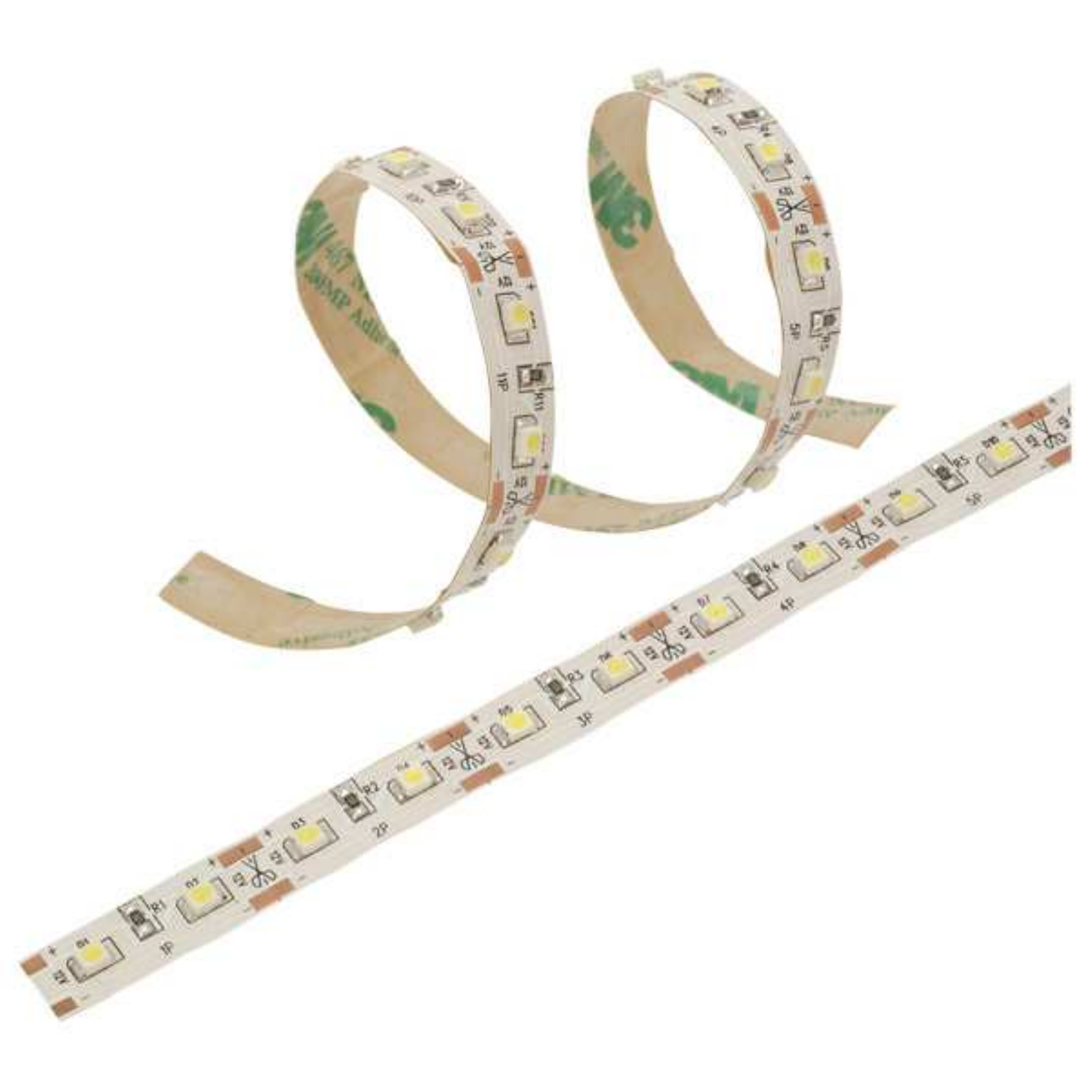}
  \includegraphics[width=0.59\textwidth]{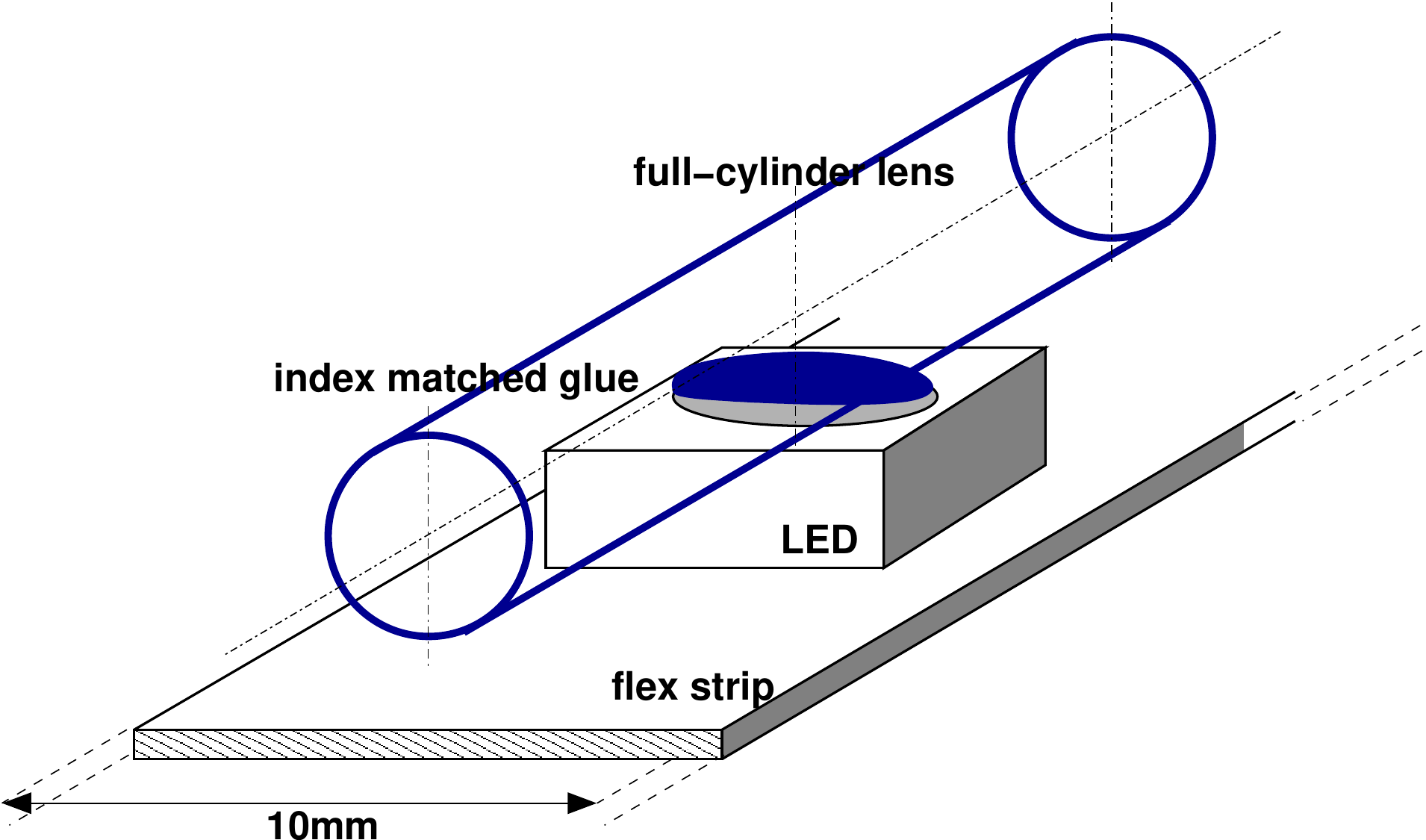}
  \caption{KingBright LED 10 mm-wide strip (left), and
  optically-coupled acrylic 
  lens (right). \label{fig:leds}}
\end{figure}

Each LED pulser drives two KingBright LSL-062-05 flexible LED lighting
modules. The LED strips are fitted with an optically coupled, cylindrical
lens, made of $2$\,mm-diameter extruded perspex, as shown in
figure \ref{fig:leds}. The KingBright LEDs feature an in-built lens
with $1/2$\,angle $120$\,$^{\circ}$ producing a rather wide beam
divergence that results in a $75\%$ loss of emission at the layers
furthest from the LED strips. The addition of the perspex rods focuses
the light so as to reduce this loss to only $25\%$ of the maximum.
\subsection{LI installation}
\label{li:installation}
The \dsecal~ JBs were installed and cabled in Aug 2010. The remainder of
the JBs and the front end electronics were installed during Oct/Nov
2010.
\begin{tabbing}
The \pecal~CC receives\hspace{11 pt}\=: $1\times$ input from the DAQ.\\
The \pecal~CC transmits             \>: $1\times$ output to each of six JBs.\\
\end{tabbing}
A single \pecal~CC, located in the LI electronics crate, transmits the
DAQ instructions and MCM clock information to the six \pecal~ JBs. 
Each JB splits and distributes the instructions and clock information to $14 =
4\times 2\; \textrm{pulsers} + 2\times 3\; \textrm{pulsers}$ forming
six groups, permitting any combination of the six \pecal~ modules to be
illuminated. 
\begin{tabbing}
Each \becal~CC receives\hspace{11 pt}\=: 1$\times$ input from the DAQ.\\
Each \becal~CC transmits             \>: 2$\times$ outputs to each of six JBs.\\
\end{tabbing}
There are six \becal~CCs in total. One \becal~CC transmits the DAQ
instructions and MCM clock information to one \becal~ JB over two RJ45/Cat 5e
cables. One JB splits and distributes the instructions and clock
information to 14 pulsers, forming three distinct groupings, each
corresponding to pulsers located on one of the three readout
faces. Any combination of readout faces (i.e. the ends of the
single and double-ended scintillator bars) can be illuminated within a
given \becal~ module. 

The \dsecal~ JBs are unique in that they sit inside the ND280 magnet,
situated directly against the \ecal~ bulkheads. Moreover they employ
LEMO connections along single core coaxial cable rather than the
RJ45/Cat 5e that is ubiquitous elsewhere
throughout the {\ecal}s and the remainder of ND280. There are two readout
faces per \dsecal~ module each separately illuminated by eight
pulsers. There are two JBs, one North and one South
served by a single CC situated in the LI front-end electronics
crate.

\section{Testbeam}
In 2009, the \dsecal~ module was exposed to the CERN T9 testbeam. This
represented the first opportunity to test the entire integrated
system: calorimeter, readout system, DAQ system and
analysis 
framework. As well as a system shakedown, the data provided by the
testbeam have been used to tune the particle identification algorithms
used in T2K analyses and to obtain a better understanding of detector
performance.

\subsection{The CERN PS T9 Testbeam}
The T9 beamline is a medium energy, multiple particle species,
tertiary beam generated by the collision of protons from the CERN PS
on a solid target in the CERN East Experimental Hall.  The dual
polarity beam supplied a mix of pions, electrons and protons
with momenta ranging from 300~MeV/c up to 15~GeV/c. Data were taken at
momentum points between 300~MeV/c and 5.0~GeV/c. The particle
composition of the beam changed with momentum setting. Below 1~GeV/c,
electrons made up 90\% of the beam. This fraction decreased to
approximately 5\% for momentum settings above 3.0~GeV/c, with a
corresponding increase in the fraction of hadrons.

\subsection{Triggering and event selection}

The beamline was instrumented with several detectors to provide
particle identification information.  Two \v{C}erenkov counters filled
with CO$_{2}$ at variable pressure were present, and read out using
PMTs connected to an ordinary TFB board. The counters were configured
such that only electrons were above threshold, which provided
electron/hadron discrimination. The electron identification efficiency
of the system (requiring both counters to register a hit) was
estimated to be $(90 \pm 2)$\% for momentum settings below
3.0~GeV/c. The \v{C}erenkov system was complemented by a time-of-flight
(TOF) system comprised of two scintillator paddles read out by fast
PMTs. The paddles were separated by a flight distance of
14 metres. Custom NIM-based electronics, including a time-to-analogue
converter (TAC), converted the time of flight to a current pulse,
which was read out using a TFB. The TOF system provided proton-pion
separation for momentum settings between 600~MeV/c and 1.8~GeV/c, and
also acted as the hardware trigger for the \dsecal.

Combining the performance of the TOF and \v{C}erenkov systems with the
beam composition, the contamination of the hadron sample with
electrons was estimated to be below 0.5\%, and the electron sample was
estimated to be more than 99\% pure. Because of the relative timings
of the physics signal and DAQ cycle, the physics data from a single
beam event could fall into one of two Trip-T integration windows. Due
to latencies in the system it was also possible that different parts
of a single event could end up spread across different time buckets.
In this case it was impossible to be sure that activity did not occur,
undetected, during the reset period. To eliminate this uncertainty,
events with reconstructed clusters in both windows were rejected. For
electron events, the candidate cluster was also required to arrive
between 0 and 65~ns after the \v{C}erenkov signal.

\subsection{Detector configuration}

The detector was configured as similarly as possible to its final
operating conditions; however, there were some differences, largely
due to physical constraints. The downstream end of the detector (with
respect to the J-PARC beam) faced upstream in the testbeam, meaning
that incident particles passed from the back to the front of the
detector. This meant that the particles passed through a lead layer
before a scintillator layer, leading to a small difference in
behaviour. 

In addition to this, the detector was cooled using an air-based
chiller rather than water cooled, leading to a diurnal variation in
temperature of around $2^{\circ}$C, and a total temperature range of
16--$28^{\circ}$C over the whole running period. This led to rather
larger drifts in MPPC behaviour than expected for the final system,
leading to some difficulties in calibration (section
\ref{testbeamCalDifferences}), though after an initial commissioning
period these were mostly ameliorated by taking very regular
calibration runs.

\subsection{Calibration differences}
\label{testbeamCalDifferences}

Calibration for the testbeam was carried out using only dedicated
pedestal runs rather than using noise spectra as in the final
calibration scheme, described in section \ref{calibration}. Initially,
pedestal runs were taken infrequently, but after examining the early
data we noticed that this was not sufficient to fully calibrate for the
effect of the large temperature variations in the T9 setup.  We
therefore changed operating procedure to take a dedicated pedestal run
before every physics run.  For operations at J-PARC, we have added the
ability to take interspersed pedestal data and measure noise spectra
during normal running, reducing the need for these dedicated runs. The
more sophisticated cooling system at J-PARC also reduces the need for
very frequent calibration.

To improve the calibration of early data, we made use of the fact that
the average temperature of the detector was measured at the start of
each run. The gain for each channel was then calculated on a
run-by-run basis, using the gains calculated from the most recent
pedestal run, the temperature measurements, and the following formula:

\begin{equation}
G = G_{0} + \frac{\textrm{d}G}{\textrm{d}T}T_{\mbox{\scriptsize{diff}}},
\end{equation}
where $G_{0}$ is the gain of the preceding pedestal run,
$T_{\textrm{diff}}$ is the temperature difference between the current
run and the preceding pedestal, and $\frac{\textrm{d}G}{\textrm{d}T}$
is the rate of change of gain with temperature, found to be
-0.67$\times 10^5/^\circ$C.
This removed much of the charge scale variation between runs.

\begin{figure}
\begin{center}
\includegraphics[width=0.98\textwidth]{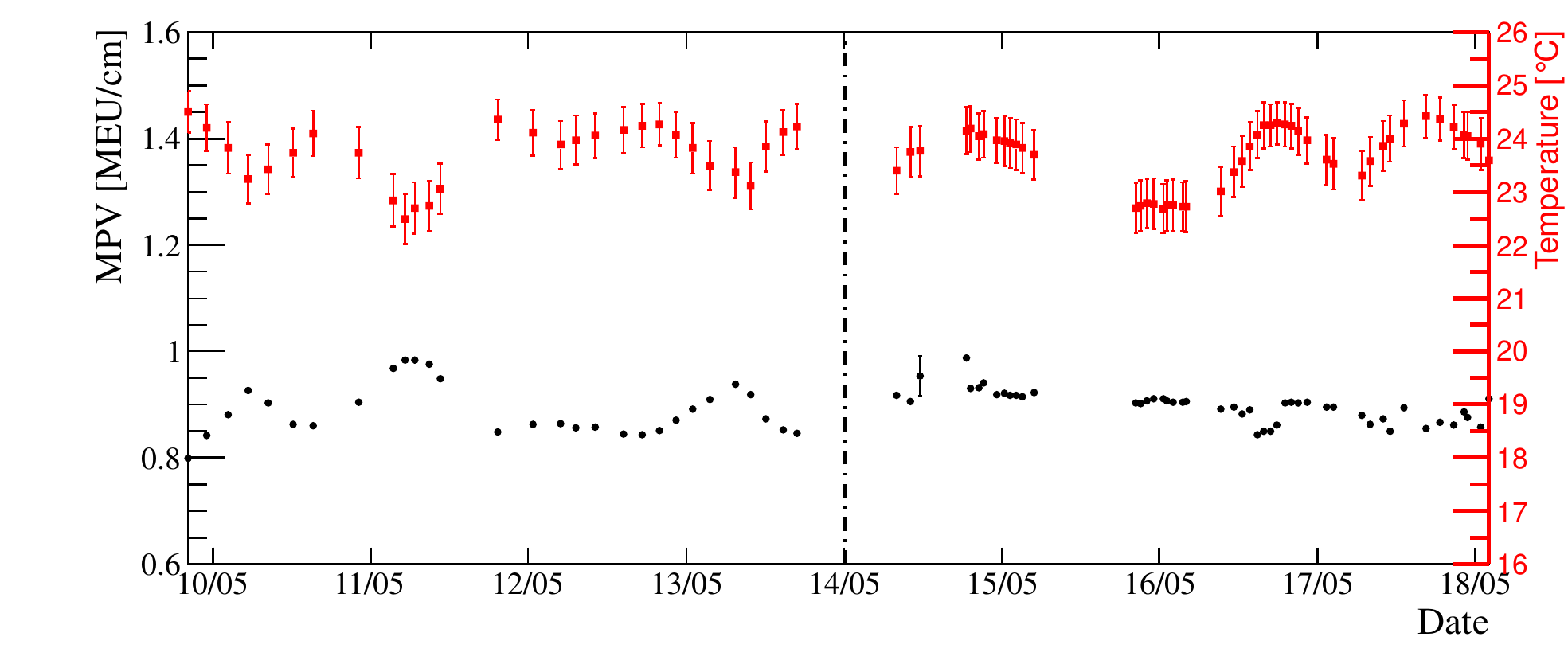}
\caption{Time series of the average detector temperature and Landau
  most probable value (MPV) for muon hits in each run, after making a
  temperature-based gain 
  correction. The dashed line represents the point at which pedestal
  runs began to be taken regularly.}
\label{tempMPV}
\end{center}
\end{figure}
As shown in figure \ref{tempMPV}, some residual temperature
dependence is seen in the detector response after making the
temperature-based correction, particularly in the period before
regular pedestal runs were taken. A fit to the
temperature data corresponding to the final five days in figure \ref{tempMPV}
indicates that the average 
temperatures were rising, resulting in the slight decrease in the Landau 
most probable values (MPVs) that is visible in the plot.  To remove
these variations, data 
from cosmic ray muons (collected coincidentally with beam events, in
DAQ integration periods not associated with the beam) were processed
from each run. The MPV of the charge
distribution of single hits across all channels was then found for
this sample, and used to scale hit charges for electron events in the
run, equalizing the MIP scale for all data runs.  

Monte Carlo samples for each individual testbeam run were produced using
Geant4 \cite{g4}. A representation of the \dsecal~ geometry was constructed, and a
monochromatic beam of electrons at each central beam momentum point was
fired into the centre of the back-face of the \dsecal. The photosensor
response was simulated utilizing data from test bench measurements and tuned
using cosmic ray muon data, with special attention to simulating the
temperature response of the photosensor. Monte Carlo samples were generated
with temperatures identical
to the average local temperature of each photosensor on a run-to-run basis.
A similar cosmic ray muon sample was also generated at this temperature, and
the calculated MPV was used to scale the electron hit charges. 

In order to remove any poorly understood noise or readout-chip threshold
variation, a charge threshold equivalent to 0.4 MIP units was implemented in
the data and simulation. In addition, during the final processing of the
data, any hits further than $8\sigma$ of the beam-time width from the
peak cluster 
time were discarded to remove unsimulated cosmic events. The Monte Carlo
data sets were then passed through the same calibration and reconstruction
paths as data.

The procedure described successfully accounted for the average gain
variation over the running period due to temperature variation. It was
not, however, possible to account for sensor-to-sensor variations or
temperature shifts within individual runs.

\subsection{Testbeam performance}

The testbeam data were used to calculate the performance of the \dsecal~
in the reconstruction of incident particles, particularly
electrons. The performance results for electrons will be shown here.

\begin{figure}
\begin{center}
\includegraphics[width=0.9\textwidth]{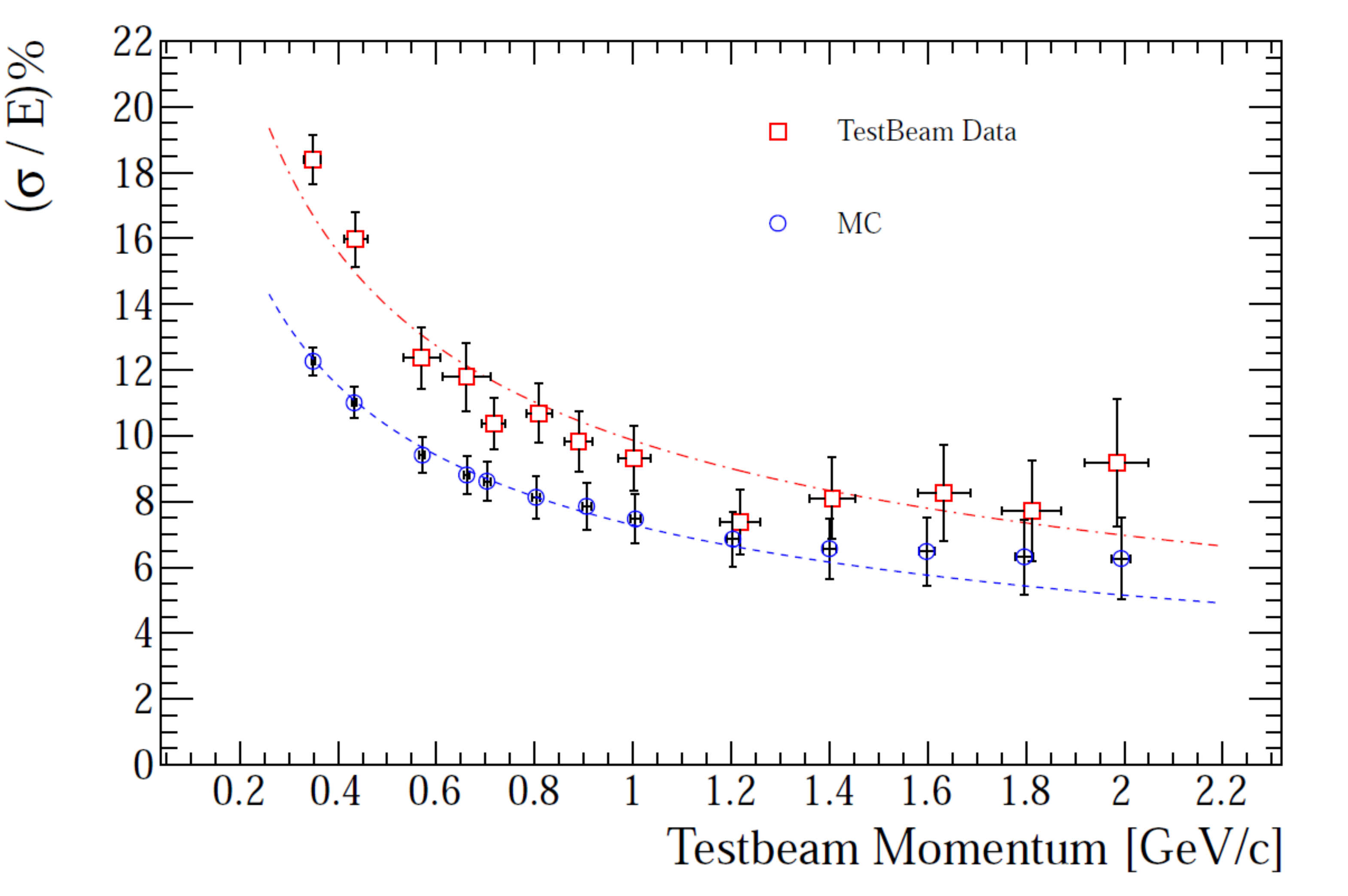}
\caption{Measured energy resolution of the \dsecal~ for electromagnetic
  showers, for data and Monte Carlo. The dashed lines show a fit to a
  stochastic resolution model.}
\label{EM_ERes}
\end{center}
\end{figure}

Figure \ref{EM_ERes} shows the measured energy resolution for
electrons striking the centre of the \ecal~ face, as a function of
energy. It can be seen that the data and Monte Carlo agree reasonably
well, but the data resolution is somewhat worse over the whole energy
range, and particularly at lower energies. This may be accounted for
in part by the intrinsic momentum spread of the beam, which was not
modelled in the Monte Carlo. In addition, the temperature-based gain
corrections used for this data will only correct the mean photosensor
gain across the \dsecal.  A spread in MPPC gains due to temperature
differences across the \ecal~ is expected, but cannot be corrected 
because the temperatures of individual sensors could not be measured
and were not modelled in the Monte Carlo, which assumes the same temperature
for all sensors. This is expected to lead to a larger spread in
measured particle energies.

Because of the limitations described above, the differences between
data and Monte Carlo seen in figure \ref{EM_ERes} represent a
worst-case scenario for the calculation of systematic uncertainties. Work is
currently underway to replace this conservative estimate with one
based on in-situ measurements.

\begin{figure}
\begin{center}
\includegraphics[width=0.8\textwidth]{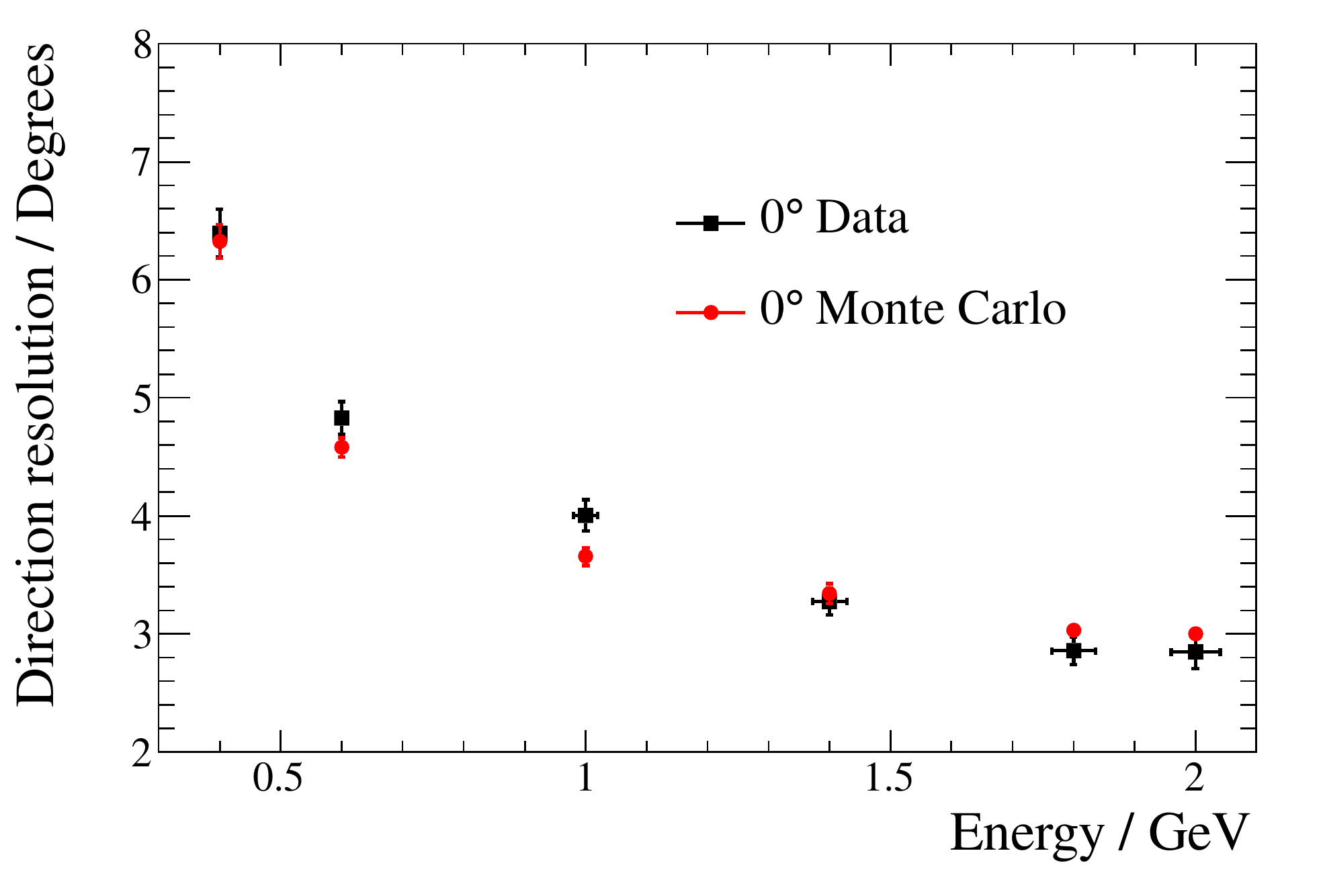}
\caption{Angular resolution for electrons at normal incidence to the
detector, in data and Monte Carlo, as a function of energy.}
\label{angRes_zeroDeg}
\end{center}
\end{figure}

Figure \ref{angRes_zeroDeg} shows the angular resolution of the
\dsecal~ for normally incident electrons at a range of
energies. Excellent agreement is seen between data and Monte Carlo
over the full range of energies considered. 

In summary, the T9 testbeam has provided useful information for the
calculation of systematics, and also gave an opportunity to test the
detector and DAQ systems before they were integrated into the
ND280. The data collected were also invaluable in the development of
calibration and reconstruction code in advance of data from the full
ND280 becoming available. 

\section{\ecal~ commissioning and performance}

\subsection{Calibration}\label{calibration}

Precise calibration of \ecal~ hits is vital for high-quality
calorimetric performance. To this end, a set of procedures has been
developed to calibrate out all significant instrumental effects, with
the goal of obtaining the energy deposited in the scintillator giving
rise to the hit, and the precise time at which the energy was deposited.
The calibration can be separated into two main categories which are
described below: energy calibration and timing calibration.

Energy calibration can be logically split into three steps---going
backwards down the readout chain from the ADC value registered by the
electronics to the anode charge from the MPPC, from this to the number
of photons incident on the MPPC face, and from this to the energy
deposit in the scintillator.

The first step involves subtracting the electronics pedestal from the
ADC counts.  This pedestal is the ADC value registered in the absence
of any physics signal, and is different for each channel and readout
cycle. Dedicated runs are taken once a week during beam down-times to
measure the pedestals separately for each cycle, and a per-channel
average over cycles is calculated every three hours from noise spectra
recorded by the DAQ in normal running.  The diurnal variation in
pedestals, due to temperature changes at the electronics, is up to a
few ADC counts, so fine-granularity corrections are important. 

\begin{figure}
\begin{center}
\includegraphics[width=0.8\textwidth]{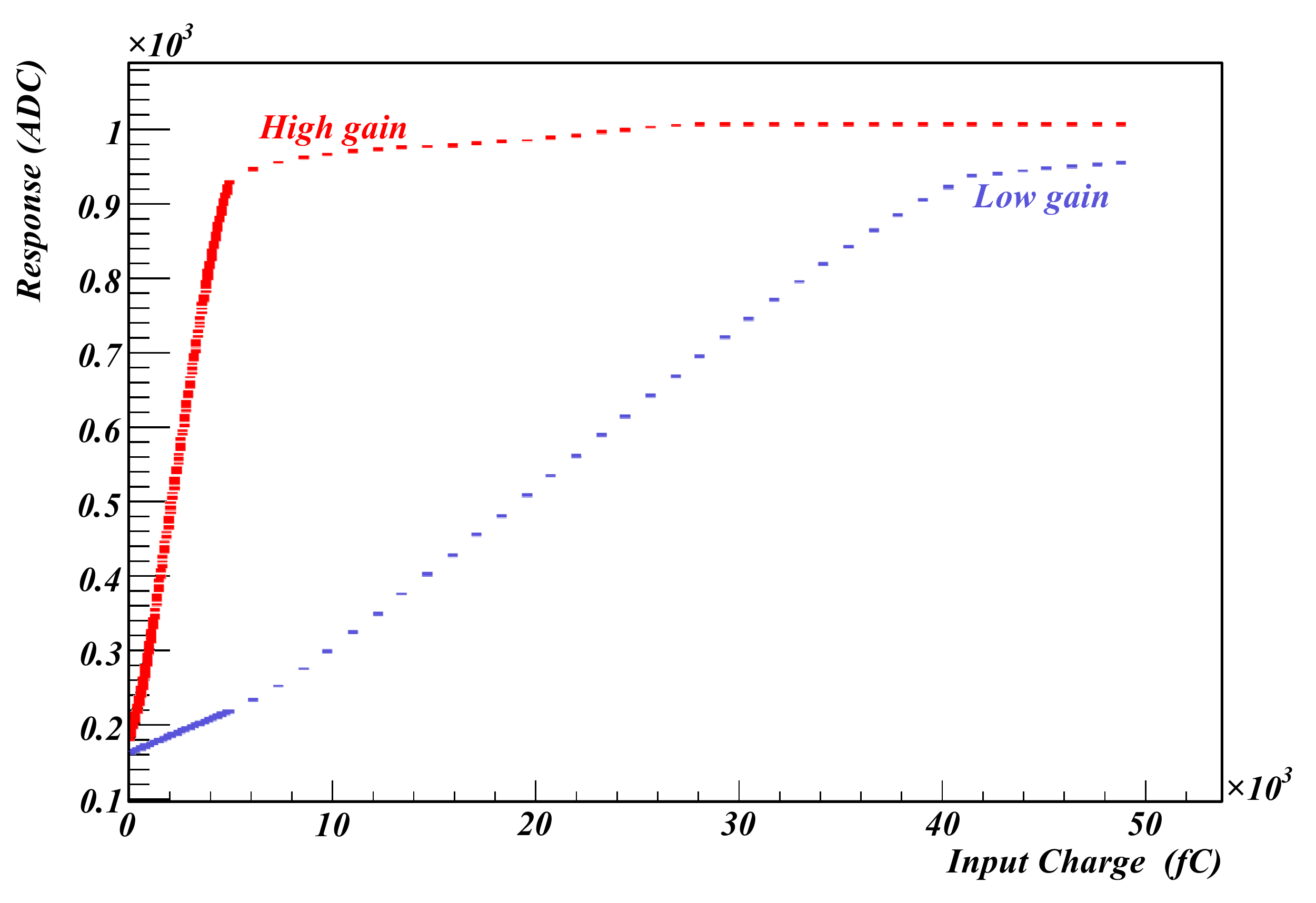}
\caption{High and low gain ADC response of a typical Trip-T channel.
 Charge is injected using an onboard capacitor controlled by a
 Digital-to-Analogue Converter (DAC). The injected charge is
 calculated from the DAC setting.  For comparison, a typical MIP
 deposits of order $10^7$ electrons or $1.6\times10^3$ fC of charge.}
\label{fig:ci}
\end{center}
\end{figure}

The next step is to convert the pedestal-subtracted ADC into an anode
charge.  The electronics response is not perfectly linear, and the
calibration must also incorporate the transition from high gain to low
gain ADC channels. Both the high gain and low gain response are mapped
using the TFB's ability to inject a known charge onto each channel in
turn.  These charge injection curves are parametrized for both high
and low gain channels using two cubic polynomials joined with a
sigmoid function to smoothly transition between them.  The
parametrized charge injection curves also 
allow us to calibrate the low gain channel response to the high gain
channel.   The switchover between the two channels is made at
approximately 460 ADC counts in the high gain channel, which typically
corresponds to about 21 PEU.  An example of the charge injection
curves is shown in figure \ref{fig:ci}.  A typical MIP deposits of
order $10^7$ electrons or 
$1.6\times10^3$ fC of charge, i.e. in the first bin on the plot.  More
information about the charge injection calibration is available in
reference  \cite{thesisWaldron}.  The
parameters for this calibration are rather stable 
over time, so are only updated around once per running period.  

To convert the calculated anode charge to an estimate for the number
of photons incident on the MPPC, we first convert the charge into a
number of PEU, by dividing by the gain of the MPPC. The gains for each
channel are calculated for every three hours of data, using the same
noise spectra as the pedestal drift, and fitting the position of the
first non-pedestal peak in the noise spectrum which is due to a single
pixel firing in the device. Again, diurnal temperature variations have
non-negligible effects on the gain, so this granularity is
required. We then need to convert this number of pixels to a number of
photons. For low light levels, this conversion can be approximated by
dividing by the PDE of the MPPC;
however, it is complicated by saturation effects (the MPPC has a
finite number of pixels and therefore a limited dynamic range), and
also by after-pulsing and crosstalk, which cause the effective PDE to
be larger than the true value \footnote{A detected photon has
  approximately a 10\% probability of initiating a secondary
  avalanche; therefore, the number of avalanches is in general higher
  than the number of detected photons}. The resulting response function cannot
easily be calculated analytically, and is therefore modelled on
testbench measurements. A single response function is used for every
channel, but the parameters are in turn functions of the device gain,
so that changes in PDE and other parameters with overvoltage can be
taken into account.

\begin{figure}[tbp]
\begin{center}
\includegraphics[width=1.06\textwidth]{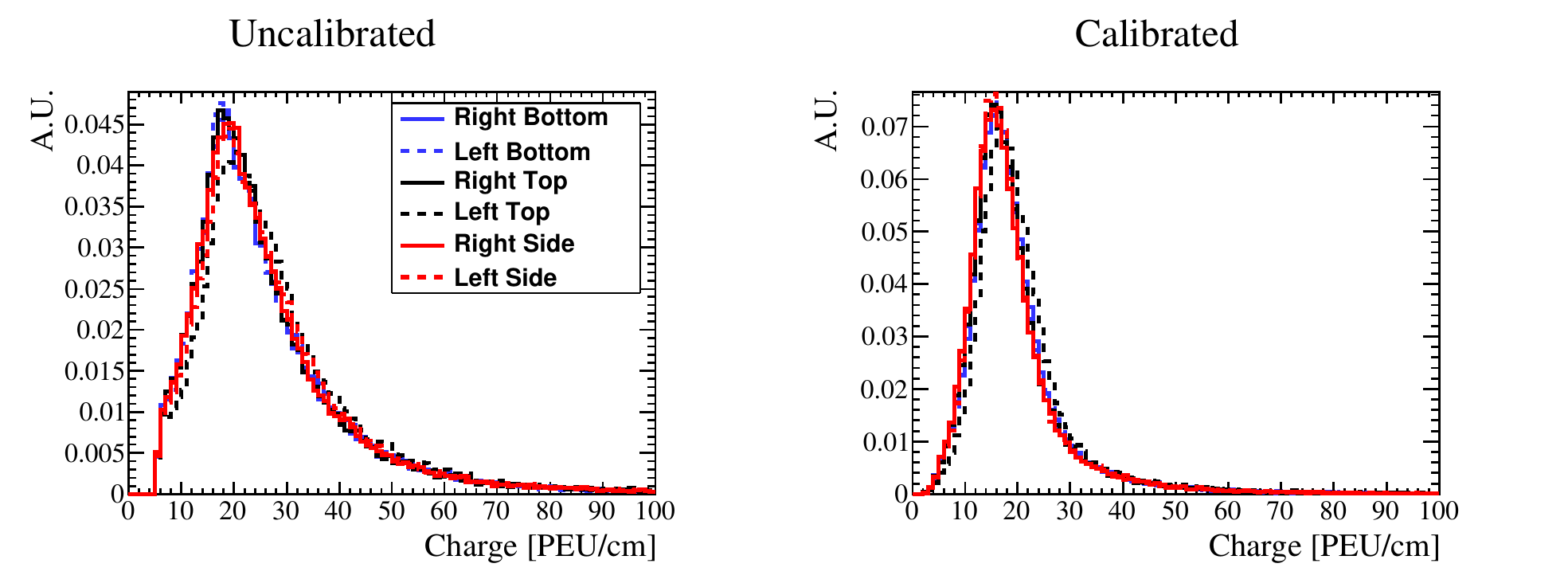}
\caption{The effect of the attenuation and uniformity corrections on
  the hit charge 
  distributions for through-going cosmic muons in the six modules of
  the \becal, shown by the six curves in each plot. The left
  side shows the distributions with the pedestal and gain corrected;
  the right side adds corrections for attenuation and differences
  caused by non-uniformity of scintillator bars and WLS
  fibres. Distributions shown here are summed within a 
  module, but the  
  corrections are applied on a per channel basis.  The slight
  variation between modules is typical and insignificant.}
\label{fig:chargecalib}
\end{center}
\end{figure}

Next, we need to convert the number of incident photons to an energy
deposit.  This needs to be done for each channel, as there will be
some non-uniformity due to, for example, the differences between
channels in the optical
coupling between the 
fibre and the MPPC. The correction is done using empirical data, taking a
sample of cosmic muon tracks passing through the \ecal, performing all
other calibrations, including attenuation in the fibre (described
below), and track path-length through the scintillator bar.  The
corrected hit size distribution for each channel is fitted with a
Landau distribution convolved with a Gaussian smearing. The effect of
the attenuation and uniformity  corrections is shown in
figure~\ref{fig:chargecalib}; 
the distributions on the right side are used for the Landau-Gaussian
fit.  This is used to calculate the expected hit size for a MIP-size
energy deposit at a fixed (1 m) distance from the sensor; the result
is used as a divisor to equalize the sizes of hits between bars.  The
signal produced by this ``ideal MIP'' varies between modules and
scintillator bar  
orientations, but to give a sense of scale, a muon passing
approximately 
perpendicularly through the centre of the \dsecal~ typically leads to
approximately 19 pixel avalanches at each sensor, corresponding to
approximately 
17 detected photons.

\begin{figure}
\begin{center}
\includegraphics[width=1.0\textwidth]{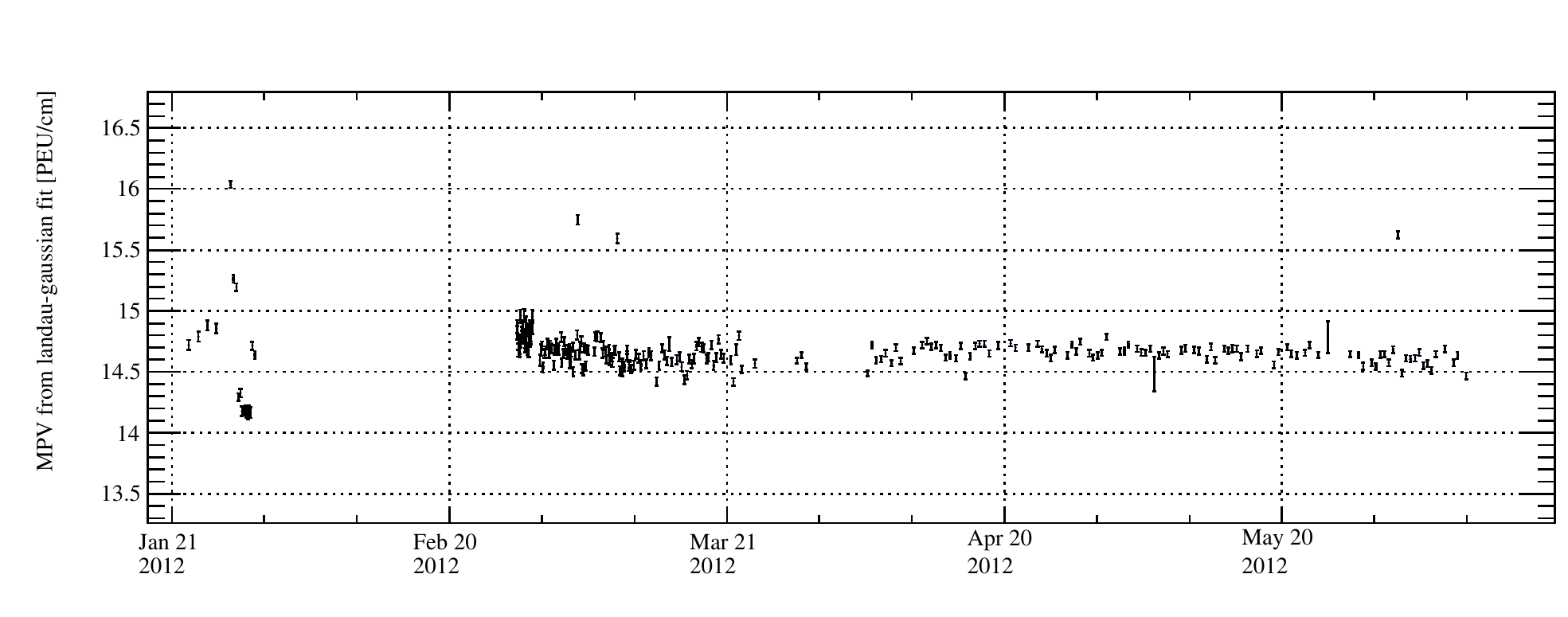}
\caption{The muon MIP response as a function of time for a period of
  beam running.}
\label{fig:MIPtimeresponse}
\end{center}
\end{figure}

It is also necessary to correct for the attenuation of photons in the
WLS fibres, which is significant over the length of the \ecal~
bars.  This calibration is done after the reconstruction has matched
tracks between the two views of the \ecal, so that the component of
the hit position along the length of the scintillator bar can be
estimated. The attenuation profile is modelled as a sum of two
exponential functions, with an additional correction for the mirroring
of the single-ended fibres. This correction is stable over time, and
is calculated separately for each orientation of bar in each module
using cosmic muon tracks. Figure~\ref{fig:MIPtimeresponse} shows the
time stability of the energy calibration over one period of beam
running.

\begin{figure}
\begin{center}
\includegraphics[scale=0.80]{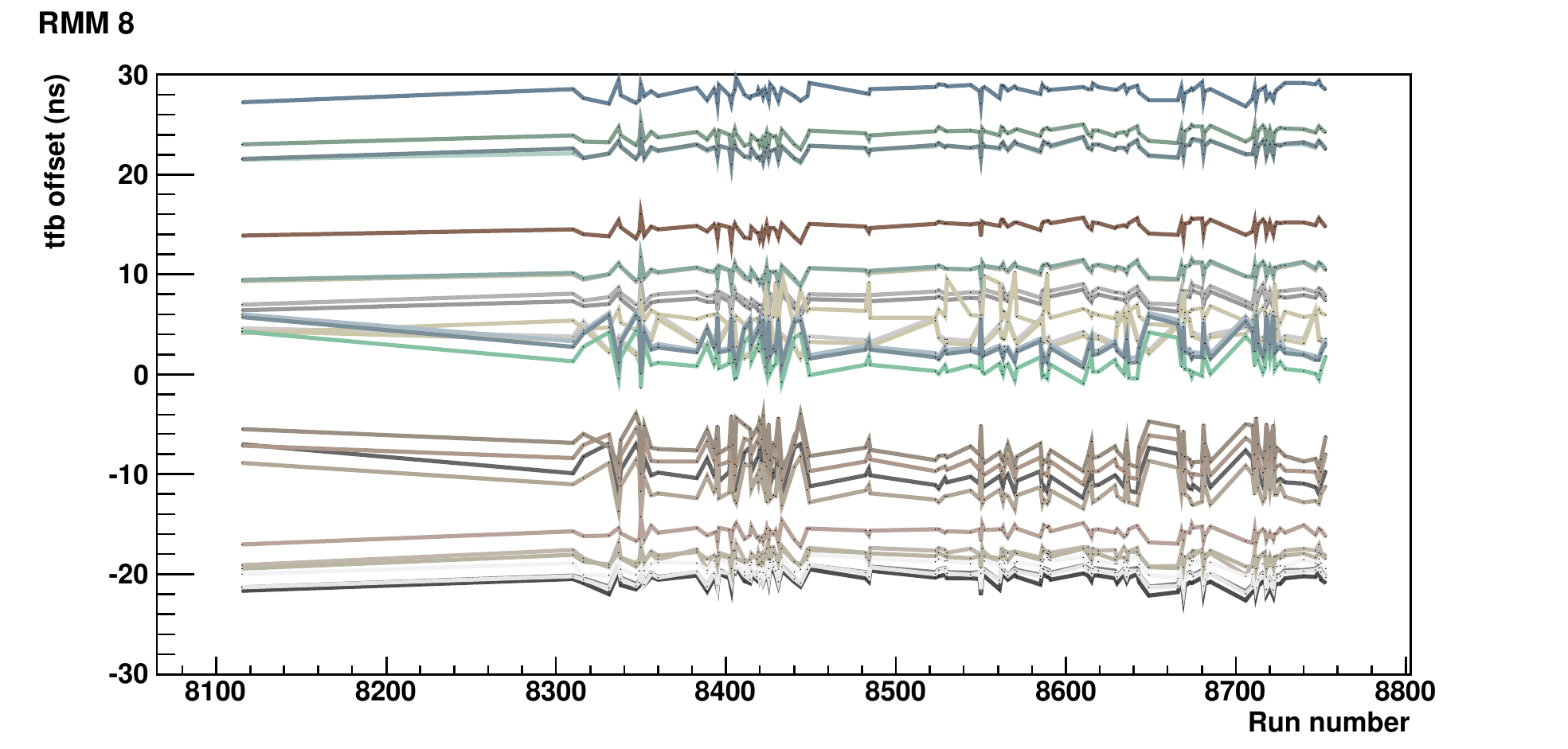}
\caption{The TFB offset before correction for all channels on one RMM as
  a function of time for a period of beam running. Each line shows one TFB.}
\label{fig:TFBoffset}
\end{center}
\end{figure}

The timing calibration also involves several distinct steps.  Each hit
is time-stamped using the TFB clock, based on the time at 
which the channel's discriminator was triggered. The discriminator
trigger is nominally set at a level of 3.5 PE.  The TFB clocks
lag the global detector clock from
which they are synchronized, due to delays in the cabling and
electronics used to distribute the clock signals. This effect will be
different for each TFB depending on the cable lengths. The differences
between TFBs within the \ecal~ are calculated by considering cosmic
muons that leave hits in channels corresponding to different TFBs,
and correcting the hit times for fibre delay and the transit time of the
muon. Synchronizing the \ecal~ with the rest of the detector is achieved
using a similar method. Figure~\ref{fig:TFBoffset} shows the offset for
the TFBs on one RMM over a period of runs.

A `timewalk' correction is also applied to the hit times.  The decay
that causes the WLS fluors to radiate occurs with a half life of a few
nanoseconds, leading to the photons arriving at the MPPC face over a
finite time period. The analogue electronics also have a finite rise time;
both of these effects result in a delay between the idealized
propagation of the signal and it crossing the discriminator threshold.
These delays depend on the total hit charge and tend to a small
constant offset for large hits.  The optical component of the timewalk
is a stochastic process and affects the timing resolution, but the
average effect can be removed with an analytic function based on an
exponential photon arrival time distribution, using the known decay
time of the relevant state in the WLS fibre.
Finally, the travel time for photons in the WLS fibre
must be calibrated. This is done after the reconstruction has
matched hits between views, so that the light travel distance is known.

\begin{figure}
\begin{center}
\includegraphics[width=0.7\textwidth]{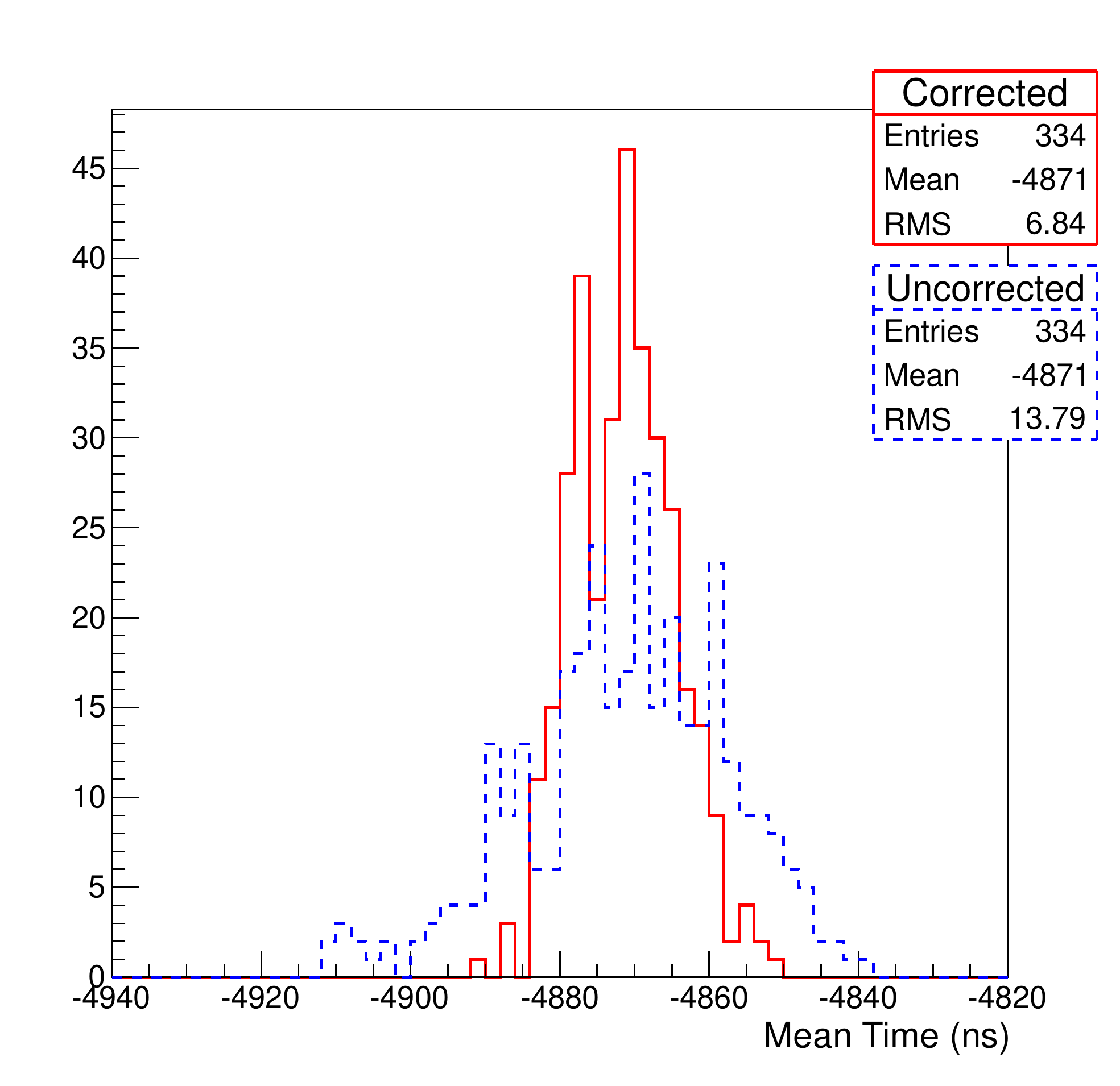}
\caption{The distributions of mean hit time values for cosmic muons
  for each channel of 
  the \tecal~  (corresponding to 334 TFBs) 
  before timing calibration (blue dashed line) and after timing calibration (red solid
line). }
\label{fig:timingcorrection}
\end{center}
\end{figure}

Figure~\ref{fig:timingcorrection} shows a comparison between the mean
hit time values of each TFB channel during cosmic muon triggers before and
after corrections. The distribution of mean hit time values narrows by a factor
of two in the RMS, and approaches the limit set by the trigger logic.
The central value corresponds to the 
time delay between a cosmic muon being observed and the corresponding
trigger being issued by the master clock. The `ideal case' resolution
for this delay is approximately 4~ns RMS, but it is dominated by the
10~ns granularity of MCM clock and the fluor decay time, both of which
are non-Gaussian.  Resolution on the time between two hits from a
single muon is rather better, and the time resolution on the entire
reconstructed \ecal~ track is better 
still.  The dominant uncertainty comes from the fluor, so each hit
provides a largely-independent estimate.  For a typical track or
shower reconstructed in the \ecal, the time resolution is approximately 1~ns or less, which is sufficient for direction discrimination when used in
combination with information from other subdetectors.

\clearpage

\subsection{Hit efficiency}

\begin{figure}
\begin{center}
\includegraphics[width=0.7\textwidth]{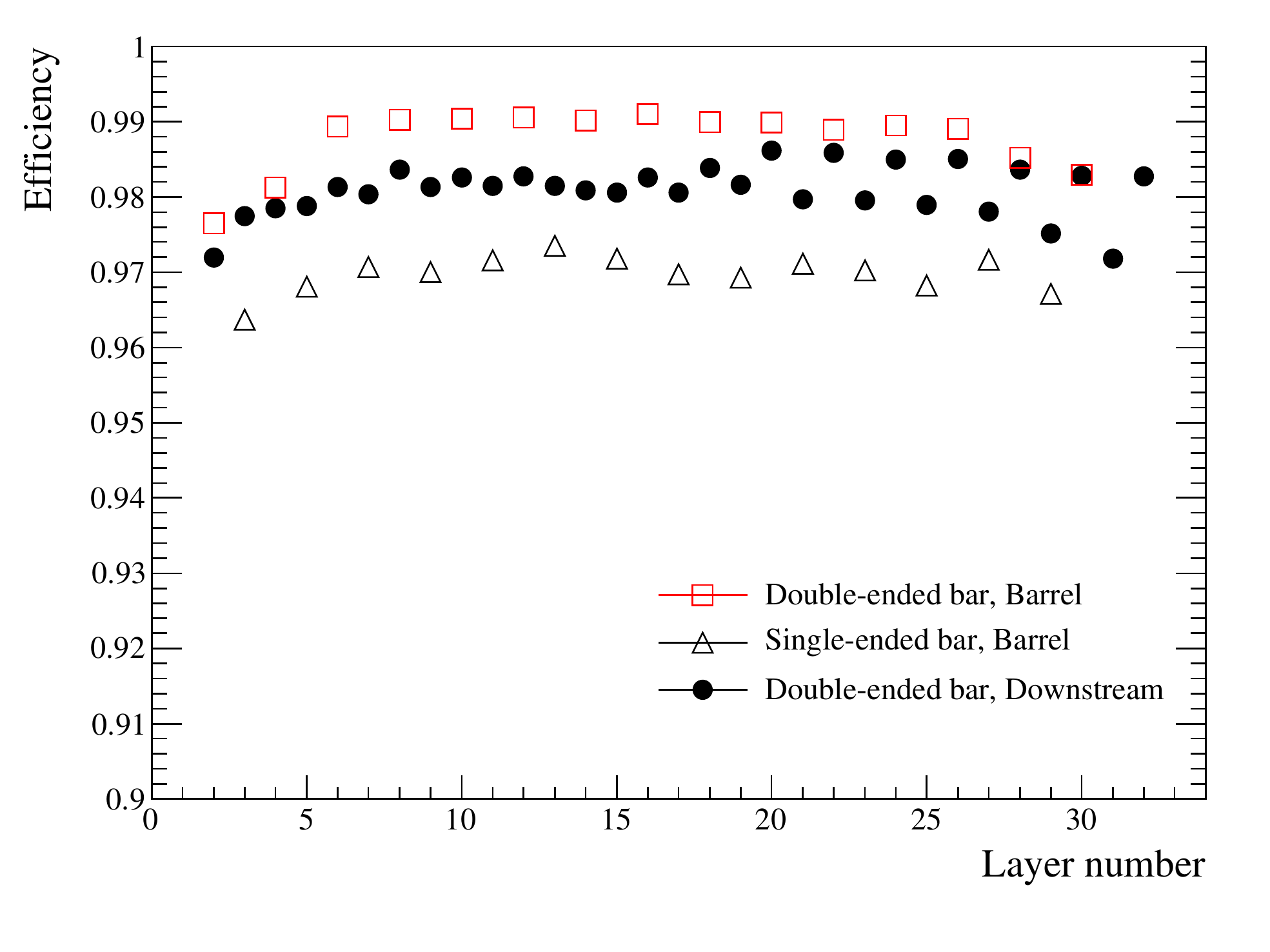}
\caption{The hit efficiency by layer for bars in the \dsecal~ and
  \becal. Layer number increases from the central region to the exterior
  of the detector.}
\label{fig:hiteff}
\end{center}
\end{figure}

The hit efficiency for the \ecal~ can be determined by looking at a
sample of through-going cosmic muons; if the scintillator bars in layer $n+1$ and layer $n-1$
are hit, the cosmic ray should have passed through a scintillator bar in
layer $n$. The sample of cosmic rays used for this measurement ensures
that the cosmic rays are isolated from other activity in the \ecal,
resulting in an accurate measure of the layer-by-layer efficiency.
The hit efficiencies by layer 
are shown in figure~\ref{fig:hiteff}. The average layer efficiency
in the \dsecal~ is 98.1\% and the average layer efficiency in the \becal~
for double-ended bars is 98.8\% and for single-ended bars is 97.0\%. The
lower efficiency in the single-ended bars is due to the reduced light
collection from only one MPPC. 

\clearpage

\subsection{Time stability and beam position}

\begin{figure}
\begin{center}
\includegraphics[width=0.9\textwidth]{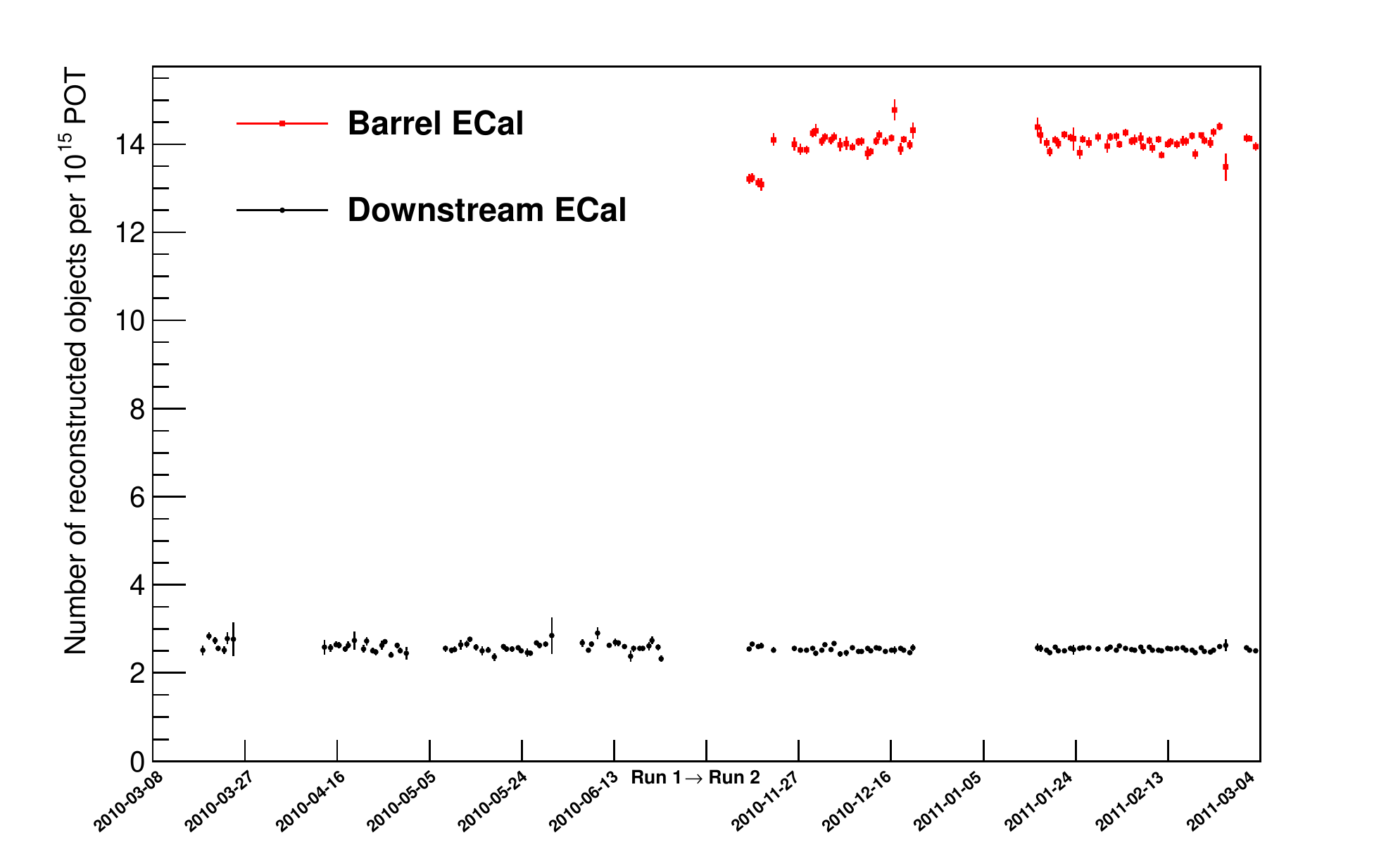} 
\caption{The rate of interactions in the \ecal~ during beam operations
  over time for T2K Run 1 and Run 2. The black points (at $\approx$ 2) show the
  \dsecal~ and the red points (at $\approx$ 14) show
  the \becal. Note the gap marked on the time axis indicating the gap
  between Runs 1 and 2, when the \becal~ was installed. }
\label{fig:beamtime}
\end{center}
\end{figure}

The \dsecal~ was in operation at the ND280 site since November 2009,
and the \becal~ since November 2010. Over
that time period, the operation of the \ecal~ has been stable. The
stability of the operation can be observed from the interactions
measured while the beam is on. A low-level selection
requiring only the reconstruction of an energy cluster 
produces a sample of events occurring in the
\ecal. Note that this selection does not correspond to a physics
selection, but simply allows for an accounting of activity in the \ecal~
while the beam is on.  Figure~\ref{fig:beamtime} shows the number of
interactions in the \ecal~ as a function of time, normalized to the
number of protons on target (POT). The rate is steady over the run periods.

\clearpage

\begin{figure}[tbp]
\begin{center}
\includegraphics[scale=0.50]{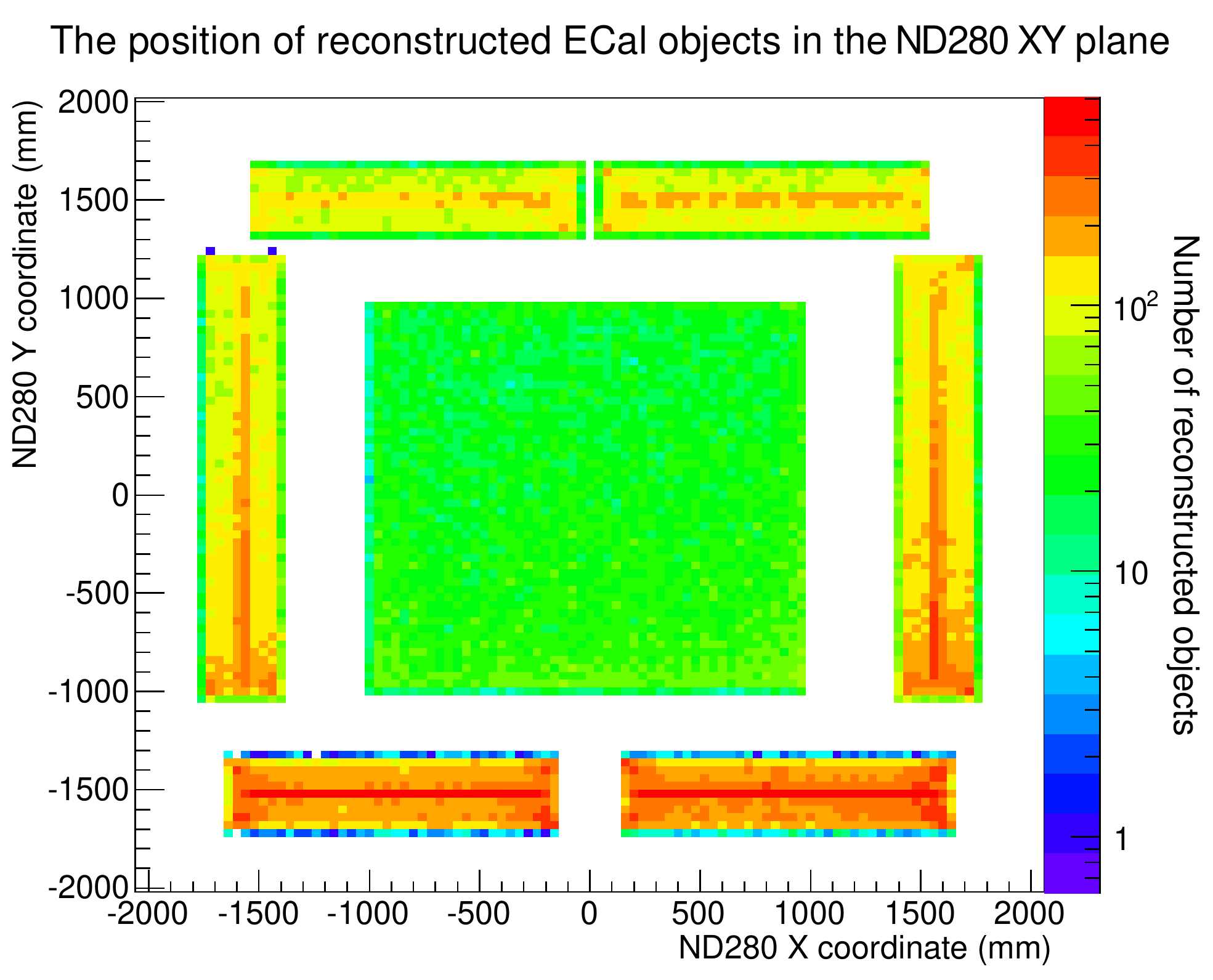}
\caption{The charge weighted position of interactions reconstructed
  in the \ecal. The position of the centre of the beam is to the lower right,
  and as a result, more events occur in the bottom right of the \ecal~
  than the top left. Events that are recorded in the centre of each
  module are through-going MIPs that leave approximately the same charge
  in each layer, and therefore have a charge weighted position in the
  centre.}
\label{fig:beampos}
\end{center}
\end{figure}

Since the ND280 detector operates off axis, the number of events as a
function of spatial position is expected to be greater near to the axis
of the beam and lesser away from the axis of the
beam. The position of the beam centre  in the coordinates of the ND280
is approximately (+3200, -9400) mm, below 
the lower right quadrant of figure~\ref{fig:beampos}, and as shown in 
the figure, the number of interactions
near the beam is significantly greater than away from the beam.  The
beam spread is approximately 4.5 m as measured by the INGRID on-axis
detector, and depends on neutrino energy.

\subsection{Particle identification}

One of the goals of the \ecal~ is to provide discrimination between
electrons and muons in the region of phase space that is not covered by
the particle identification from the tracker. The first method for
particle identification constructed for the \ecal~ was intended to
maximize the separation of track-like (muon-like) and shower-like
(electron-like) interactions. The discriminator is constructed as a
neural network~\cite{bishop1995neural} using low level quantities from the
reconstruction.

To test the discriminator, samples of muons and electrons are selected
from the data. For the \dsecal~ (\becal), the muons selected are
`through-going' 
muons that have one track component in each of the three TPC
modules and the \dsecal~ (\becal). Additionally, the track must appear
`muon-like'  
to the TPC~\cite{tpc}.   The electron sample is
produced by looking at photon pair production in the 
FGD~\cite{fgdNIM}, requiring tracks with opposite charge, at least one
TPC component in the 
track, and that the TPC component appears `electron-like'. Electrons or
positrons that enter the \tecal~ are added to the sample.  In these
control samples, the median muon momentum is 1.7 GeV/c, with muon
momenta ranging from 300 MeV/c to 10 GeV/c; the median
electron or positron momentum is 165 MeV/c.  These energies are
typical of events seen from the 
neutrino beam. Additionally, the discriminator is insensitive to the muon
momentum provided that it is greater than approximately 300 MeV/c; that is,
provided that the muon is MIP-like.

\begin{figure}[p]
\begin{center}
\includegraphics[scale=0.50]{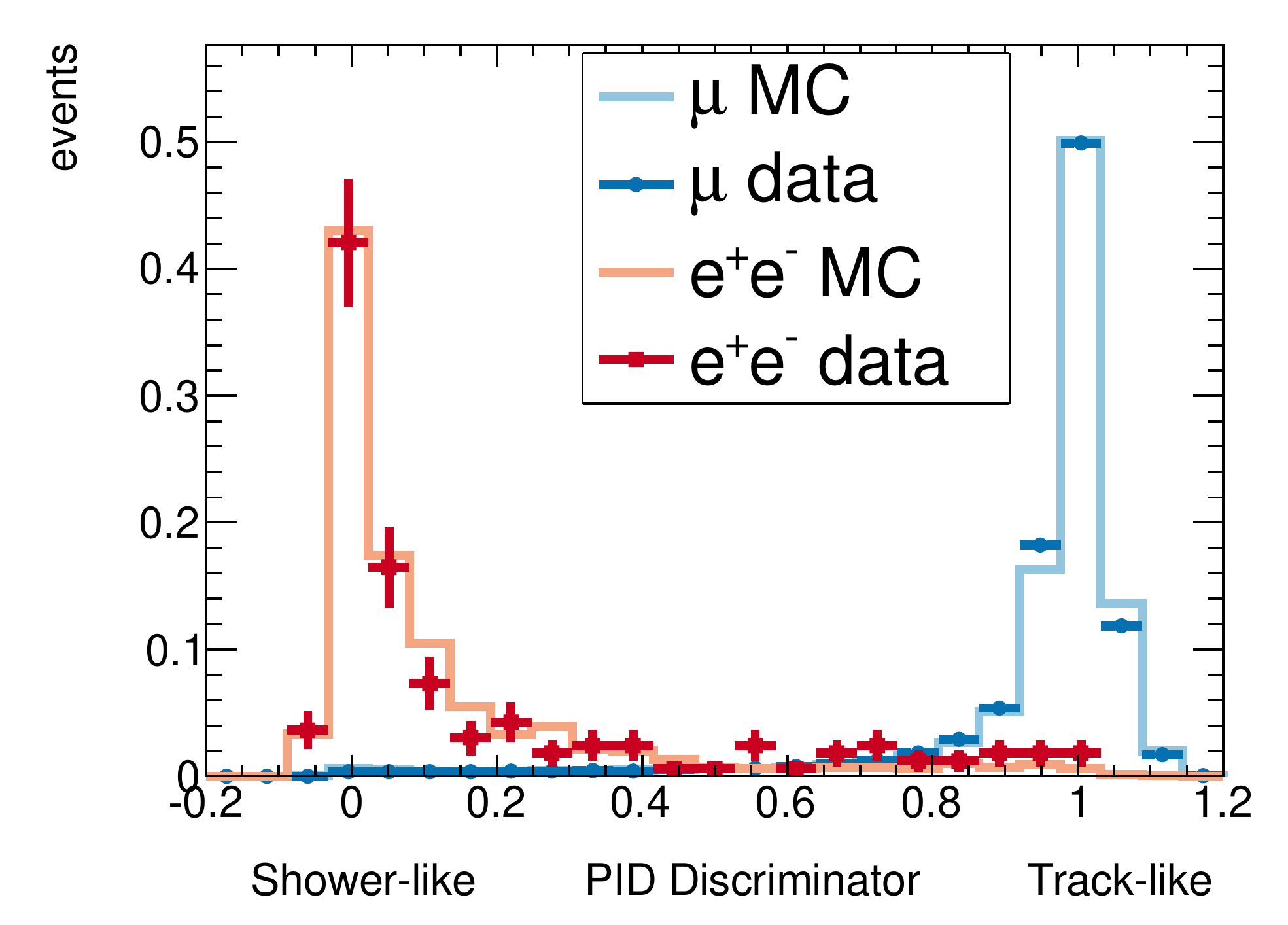}
\caption{The discrimination between track-like (muon-like) and
  shower-like (electron-like) samples in the \dsecal. Solid lines show
  Monte Carlo information and points show data samples described in the text.}
\label{fig:trshds}
\end{center}
\end{figure}

\begin{figure}[p]
\begin{center}
\includegraphics[scale=0.50]{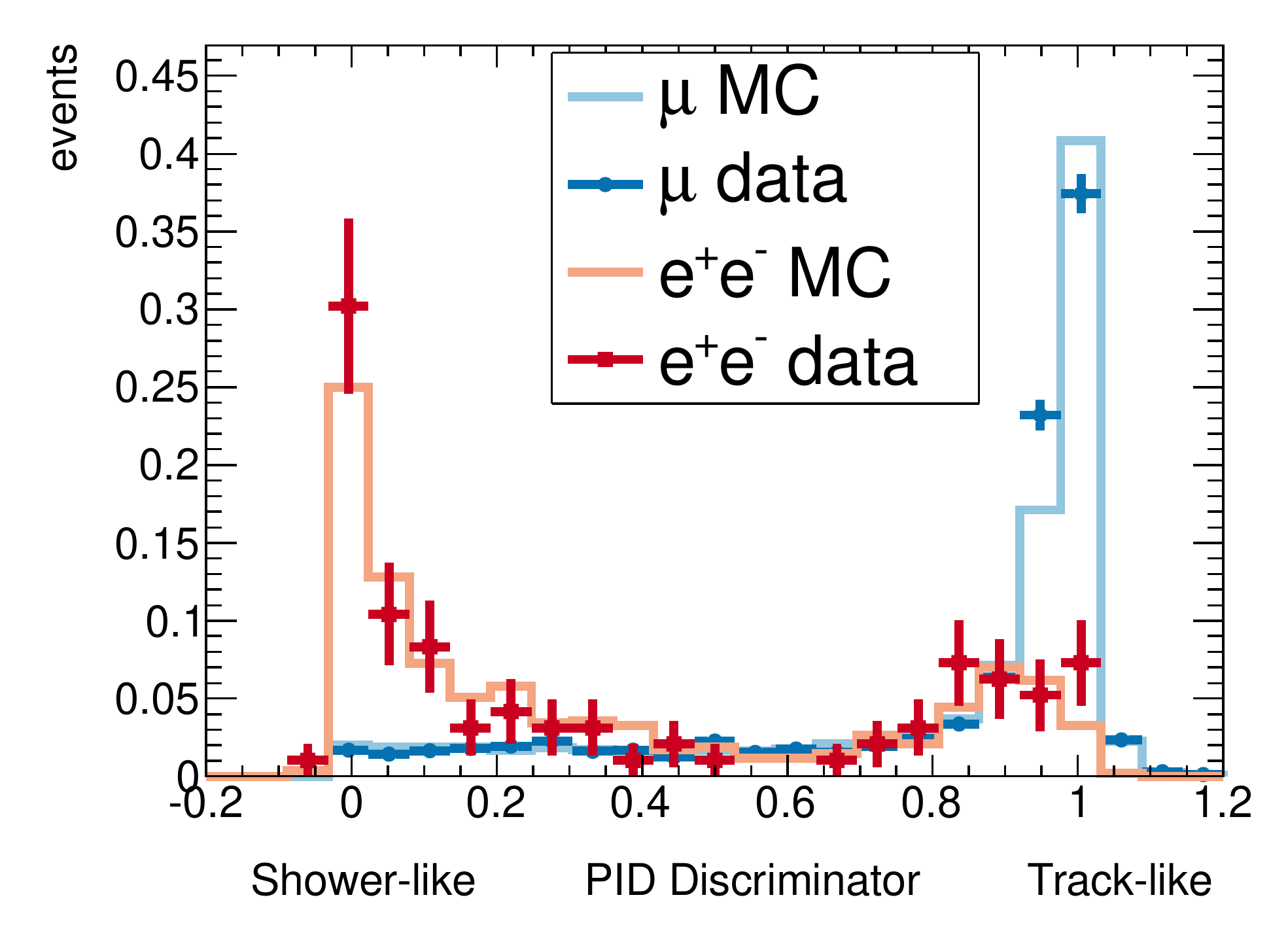}
\caption{The discrimination between track-like (muon-like) and
  shower-like (electron-like) samples in the \becal. Solid lines show
  Monte Carlo information and points show data samples described in the text.}
\label{fig:trshbar}
\end{center}
\end{figure}

The electron and muon samples are shown with the calculated
discriminator in figures~\ref{fig:trshds} and~\ref{fig:trshbar} for
the \dsecal~ and \becal, respectively. For both, there is good agreement
between the data and the Monte Carlo, and good separation between the
muon and electron samples; however, the separation is somewhat better
in the \dsecal~ than in the \becal~ due to the different $p-\theta$
distributions seen in each.  The \becal, because of its position
within the ND280, sees, on average, lower momentum particles with a
higher angle of incidence than the \dsecal.  This combination of
properties presents greater challenges for reconstruction and particle
identification.

\section{Summary}

The electromagnetic calorimeter (\ecal ) for the T2K ND280 was
designed and constructed in the UK during the period 2007-2010. The first
module, the \dsecal, was tested in a charged-particle beam at CERN in
spring 2009 and installed in the ND280 at J-PARC in
time for the first T2K neutrino beam in January 2010. The testbeam
data validated the design and operation of the system and have been
used to characterize its response. The remaining 12 \ecal~ modules were
constructed in 2009-2010 and installed at J-PARC in time for
the second neutrino data-taking period of T2K starting in fall
2010. The \ecal~ has been operating stably and survived the great
earthquake of March 2011 without any damage. The \ecal~ has met its
goals in terms of energy resolution and particle identification and is
an integral part of the ND280. \ecal~ data are used in the
ND280 physics analyses and this is becoming more important as
larger data samples are collected and more sophisticated analyses are
being developed.

\clearpage

\acknowledgments
The \ecal~ detector has been built and is operated using funds
provided by the Science and Technology Facilities Council
UK. Important support was also 
provided by the collaborating institutions. Individual researchers have been
supported by the Royal Society and the European Research Council. The
authors also wish to thank CERN for support with the testbeam, FNAL for
manufacturing the scintillator bars and preparing the wavelength-shifting
fibres, and finally our T2K colleagues for their invaluable help during
installation and commissioning of the detector.

\newcommand{\arxiv}[2]{\href{http://arxiv.org/abs/arXiv:#1}{\tt arXiv:#1\,[#2]}}
\newcommand{\arxivnosec}[1]{\href{http://arxiv.org/abs/arXiv:#1}{\tt arXiv:#1}}
\bibliographystyle{t2kukjinst}
\bibliography{manuscript}

\end{document}